\documentclass[a4paper,11pt]{article}
\pdfoutput=1 % if your are submitting a pdflatex (i.e. if you have
\usepackage{jheppub} % for details on the use of the package, please
\usepackage[T1]{fontenc} % if needed

\usepackage{slashed}
\usepackage{graphics,bm}
\usepackage{epsfig}
\usepackage{graphicx}
\usepackage{amsmath}
\usepackage{amssymb}
\usepackage{bbold}
\usepackage{wrapfig}
\usepackage{color}
\newcommand{\beq}{\begin{equation}}
\newcommand{\eeq}{\end{equation}}
\newcommand{\bqa}{\begin{eqnarray}}
\newcommand{\eqa}{\end{eqnarray}}

\def\square{\vcenter{\vbox{\hrule height.4pt
          \hbox{\vrule width.4pt height4pt
          \kern4pt\vrule width.3pt}\hrule height.4pt}}}

\title{Pion and kaon condensation at zero temperature in 
  three-flavor $\chi$PT at nonzero isospin and strange chemical potentials
  at next-to-leading order
}
\author[a,b,1]{Prabal Adhikari,\note{Corresponding author.}}
\author[b,c]{Jens O. Andersen}
\affiliation[a]{Wellesley College, Department of Physics, 106 Central Street,
  Wellesley, MA 02481, United States}
\affiliation[b]{Department of Physics, Faculty of Natural Sciences, NTNU, 
Norwegian University of Science and Technology, H{\o}gskoleringen 5,
N-7491 Trondheim, Norway}
\affiliation[c]{Niels Bohr International Academy, Blegdamsvej 17,
DK-2100 Copenhan, Denmark}
\emailAdd{pa100@wellesley.edu}
\emailAdd{andersen@tf.phys.ntnu.no}
\abstract{We consider three-flavor chiral perturbation theory ($\chi$PT) at zero
  temperature and nonzero isospin ($\mu_{I}$) and strange ($\mu_{S}$) chemical
  potentials. The effective potential is calculated to next-to-leading order
  (NLO) in the $\pi^{\pm}$-condensed phase, the $K^{\pm}$-condensed phase,
  and the $K^0/\bar{K}^0$-condensed phase. It is
  shown that the transitions from the vacuum phase to these
  phases are second order and take place when,
  $|\mu_I|=m_{\pi}$, $|{1\over2}\mu_I+\mu_S|=m_K$, and
  $|-{1\over2}\mu_I+\mu_S|=m_K$, respectively at
  tree level and remains unchanged at NLO. The transition between the two
  condensed phases is first order. The effective potential in the
  pion-condensed phase is independent of $\mu_S$ and in the
  kaon-condensed phases, it only depends on the combinations
  $\pm{1\over2}\mu_I+\mu_S$ and not separately on $\mu_I$ and $\mu_S$.
  We calculate the pressure, isospin density
  and the equation of state in the pion-condensed phase and 
  compare our results with recent $(2+1)$-flavor lattice QCD data.
  We find that the three-flavor $\chi$PT results are in good agreement
  with lattice QCD for $\mu_I<200$ MeV, however for larger values
  $\chi$PT produces values for observables that are consistently above
  lattice results. For $\mu_I>200$ MeV, the two-flavor results are in
  better agreement with lattice data.
  Finally, we consider the observables in the limit
  of very heavy $s$-quarks, where they reduce to their
  two-flavor counterparts with renormalized couplings.
  The disagreement between the predictions of two and three flavor
  $\chi$PT can largely be explained by the differences in the
  experimental values of the low-energy constants.
  
}
\begin{document} 
\maketitle
\flushbottom

\section{Introduction}
Quantum chromodynamics (QCD), the theory of the strong force, is challenging to
study due to its non-perturbative nature and the inability to use lattice QCD
simulations in the phenomenologically most interesting regime, namely finite
baryon density, due to the infamous fermion sign problem~\cite{fukurev, simon}
present in classical Monte Carlo algorithms. As such, except for
asymptotically large baryon chemical potentials, where QCD is expected to be in
a color-flavor-locked phase~\cite{raja, alford} and can be studied due to
asymptotic freedom, %which allows for the construction of an effective field
%theory,
most of the phenomenologically relevant QCD phase diagram must be
mapped out by other methods, e.g. low-energy effective models.

Recently, there has been renewed interest in a slightly different regime of
QCD, one with a finite isospin chemical potential due to the possibility of a
new form of compact stars known as pion stars,
first discussed in Ref.~\cite{carigchpt}.
This type of compact object could form
in regions with large densities of neutrinos,
which in turn leads to the production of pions and their subsequent
condensation~\cite{brauner}. These pions
under weak equilibrium lead to stable pion stars, which may be
electromagnetically neutralized by either electrons or muons, or both.
They
are expected to have radii and masses that are substantially larger than those
of neutron stars%and a mass-radius relationship that is extremelystiff
~\cite{endro}. Pion stars are also different from neutron stars
in the sense that at $T=0$ it is interactions that give rise to an (effective)
equation of state, and not the statistics of its constituents.

QCD at finite isospin chemical potential was first studied by Son and Stephanov
using chiral perturbation theory ($\chi$PT)~\cite{wein,gasser1,gasser2,bein,scherer} in their seminal paper~\cite{son}.
In Refs.~\cite{loewe,fragaiso,cohen2,janssen,carig,carigchpt,luca} one can find
various applications of $\chi$PT including some partial next-to-leading
order results. Since then finite isospin systems have been studied
extensively in other versions of QCD including two-color and adjoint
QCD~\cite{cotter,kim}, in the NJL~\cite{toublannjl,he2f,heman2,heman,ebert1,ebert2,sun,lars,2fabuki,heman3,he3f,ricardo,ruggi}, in the 
quark-meson model~\cite{lorenz,ueda,qmstiele,allofus}, but also through lattice
QCD, where it
does not suffer from the fermion sign problem (except at finite magnetic
fields~\cite{endromag,prabal} due to the charge asymmetry of the up and down
quarks). The first lattice QCD calculations of finite isospin QCD were done in
Refs.~\cite{kogut1,kogut2} and a more recent, thorough analysis in
Refs.~\cite{gergy1,gergy2,gergy3}. They find as expected from chiral
perturbation
theory calculations that at zero temperature there is a second order phase
transition at an isospin chemical potential,
$|\mu_{I}|=m_{\pi}$~\footnote{The $|\mu_I|={1\over2}m_{\pi}$
conventions is also frequently found in the literature. See Eq.~(\ref{muidef}).}, which
remains largely unaltered at finite temperatures up
to approximately $170\ {\rm MeV}$ beyond which quarks become
deconfined~\cite{fragaiso}. Similarly, with increasing isospin chemical
potentials the quarks in the pions become more loosely bound and occur in a
BCS phase though owing to the fact that this phase has the same order
parameter as the BEC phase, there is no real phase transition, only a
crossover transition, with the size of the pion condensate decreasing
substantially within a narrow isospin window.

There have been a number of studies in recent years comparing $(2+1)$ flavor
lattice QCD results with both QCD models and effective theories. Recently, the
NJL model (non-renormalizable) comparisons~\cite{ricardo} were made that showed
good agreement with the lattice while the quark-meson model~\cite{allofus}
(which is renormalizable) largely agrees with the lattice. Furthermore, there
have been other comparisons of lattice QCD with results from an effective field
theory (and model-independent) description~\cite{carig}, which is valid for
asymptotically large isospin chemical potentials~\cite{cohen2}, where the pions
behave as a free Bose gas. A recent review can be found in Ref.~\cite{massrev}.

The focus of this work is to compare the results of three-flavor $\chi$PT at
finite isospin density~\cite{kogut3} with that of $(2+1)$-flavor lattice QCD of
Refs.~\cite{gergy1,gergy2,gergy3}. We previously studied two-flavor $\chi$PT at
next-to-leading order (NLO)~\cite{us} and found that the NLO results are
in better agreement with lattice QCD than the tree-level results though the
pressure, isospin density and energy density were all found to be consistently
smaller than lattice QCD values. This is not entirely unexpected since the
lattice QCD observables included the effects of the sea strange
quarks~\cite{usagain} while two-flavor $\chi$PT does not. As such, we extend
our previous work in NLO two-flavor $\chi$PT to include the effect of the
strange quarks by using three-flavor $\chi$PT at finite isospin chemical
potential and find that the observables near the second phase transition is in
good agreement with lattice QCD. As a natural extension of our finite
isospin study, we also construct the NLO, one-loop effective potential to study
the effects of the simultaneous presence of both the isospin and strange quark
chemical potential.~\footnote{Note that the ``strange quark chemical potential"
  ($\mu_{s}$) is different from the ``strange chemical potential" ($\mu_{S}$).
  We
  define them in Eq.~(\ref{musdef}).}
We find the second-order phase transition in the pion condensed phase remains
at $|\mu_{I}|=m_{\pi}$ even with the inclusion of $\mu_{S}$ and
NLO corrections.~\footnote{This property is expected to hold to all orders in
perturbation theory.}
Similarly, the second order phase transition in the kaon condensed phases remain
at $|\pm\frac{1}{2}\mu_{I}+\mu_{S}|=m_{K}$
where $m_{K}$ is the kaon mass. Furthermore
the effective potential even in the presence of $\mu_{S}$ in the pion condensed
phase only depends on $\mu_{I}$ and in the kaon condensed phase on the
combination $|\pm\frac{1}{2}\mu_{I}+\mu_{S}|$ but not $\mu_{I}$ and $\mu_{S}$
separately. 

The paper is organized as follows.
In the next section, we discuss the Lagrangian of
three-flavor chiral perturbation
theory at finite isospin and strange chemical potentials
at next-to-leading order in the low-energy expansion.
In Sec.~3, we review the ground state of the theory and fluctuations
in the different phases.
In Sec.~4 the NLO effective potential in the
three different phases of the theory is calculated. In Sec.~5, 
we calculate the pressure, the isospin density, and the equation of state
in the pion-condensed phase.
We also consider the large-$m_s$ limit, where it is shown that the
observables in three-flavor $\chi$PT reduce to the two-flavor observables
of Ref.~\cite{us} with renormalized couplings.
In Sec.~6, we  discuss 
the phase diagram in more detail and
derive medium-dependent masses at tree level.
We compare our results for the thermodynamic functions 
with recent lattice simulations.
%In the Appendices, we list the expressions for $m_{\pi}$, $m_K$,
%$f_{\pi}$, and $f_K$ as functions of the parameters and couplings
%of $\chi$PT since we need

\section{$\chi$PT Lagrangian at ${\cal O}(p^4)$}
In this section, we briefly discuss the symmetries of three-flavor QCD
as well 
the chiral Lagrangian to next-to-leading order in the low-energy
expansion and its renormalization.
The three-flavor Lagrangian of QCD is
\bqa
{\cal L}&=&
\bar{\psi}\left(
i/\!\!\!\!D-m
\right)\psi-{1\over4}F_{\mu\nu}^aF^{\mu\nu a}\;,
\eqa
where $m={\rm diag}(m_u,m_d,m_s)$ is the quark mass
matrix, $/\!\!\!\!D=\gamma^{\mu}\left(\partial_{\mu}-ig\tfrac{\lambda^a}{2}A_{\mu}^a\right)$
is the covariant derivative, $\lambda^a$ are the Gell-Mann matrices,
$g$ is the strong coupling, $A_{\mu}^a$ is the gauge field, 
and $F_{\mu\nu}^a$ is the field-strength tensor.
The global symmetry of massless three-flavor QCD is
$SU(3)_L\times SU(3)_R\times U(1)_B$, which is spontaneously broken
down to $SU(3)_V\times U(1)_B$ in the vacuum.
For two degenerate light quarks, i.e. in the isospin limit the symmetry is
$SU(2)_I\times U(1)_Y\times U(1)_B$,  where $Y$ represents hypercharge.
If $m_u\neq m_d$, this symmetry is reduced to
$U(1)_{I_3}\times U(1)_Y\times U(1)_B$.
If we add a chemical potential
for each of the quarks, the symmetry is
$U(1)_{I_3}\times U(1)_Y\times U(1)_B$, irrespective of the quark masses.

In the present paper, we consider three-flavor QCD with
two degenerate light quarks. The chiral Lagrangian then describes
the octet of pseudo-Goldstone bosons consisting of the three pions
($\pi^{\pm}$ and $\pi^0$), the four kaons ($K^{\pm}$, $K^0$ and $\bar{K}^{0}$), and the eta ($\eta$).
We begin with the chiral perturbation theory Lagrangian 
at
$\mathcal{O}(p^{2})$~\cite{gasser1}~\footnote{One factor of $\nabla_{\mu}$ 
  counts one power of $p$ and one factor of $\chi$ counts two powers
  of $p$.}
\bqa
\mathcal{L}_{2}=\frac{f^{2}}{4}{\rm Tr}
  \left [\nabla_{\mu} \Sigma^{\dagger} \nabla^{\mu}\Sigma 
    \right ]
+{f^2\over4}{\rm Tr}
\left [\chi^{\dagger}\Sigma+\chi\Sigma^{\dagger}\right ]\; ,
\label{lag0}
\eqa
where $f$ is the bare pion decay constant,
$\chi=2B_0M$, and
\bqa
M={\rm diag}(m_u,m_d,m_s)
\eqa
is the quark mass matrix, $\Sigma=U\Sigma_0U$, where
$U=\exp{i\lambda_i\phi_i\over2f}$, and $\Sigma=\mathbb{1}$ is the vacuum.
Moreover, $\lambda_i$  ($i=1,2,...,8$) are the Gell-Mann matrices that satisfy
${\rm Tr}\lambda_i\lambda_j=2\delta_{ij}$ and $\phi_i$ are the
fields that parametrize the Goldstone manifold.
The covariant derivative at nonzero quark chemical
$\mu_q$ 
potentials $(q=u,d,s)$ is defined as follows
\bqa
\label{nabla0}
\nabla_{\mu} \Sigma&\equiv&
\partial_{\mu}\Sigma-i\left [v_{\mu},\Sigma \right]\;,\\ 
\nabla_{\mu} \Sigma^{\dagger}&=&
\partial_{\mu}\Sigma^{\dagger}-i [v_{\mu},\Sigma^{\dagger} ]
\;,
\label{nablas}
\eqa
where 
\bqa
v_{\mu}&=&\delta_{\mu0}
{\rm diag}(\mu_u,\mu_d,\mu_s)\;.
\eqa
We can also express $v_{\mu}$ in terms of the baryon, isospin and
strangeness chemical potentials $\mu_B$, $\mu_I$, and $\mu_S$ as
\bqa
v_{\mu}&=&\delta_{\mu0}{\rm diag}(\mbox{$1\over3$}\mu_B
+\mbox{$1\over2$}\mu_I,\mbox{$1\over3$}\mu_B-\mbox{$1\over2$}\mu_I,
\mbox{$1\over3$}\mu_B-\mu_S)\;.
\label{vmu}
\eqa
where
\bqa
\mu_B&=&{3\over2}(\mu_u+\mu_d)\;,\\
\label{muidef}
\mu_I&=&\mu_u-\mu_d\;,\\
\mu_S&=&{1\over2}(\mu_u+\mu_d-2\mu_s)\;.
\label{musdef}
\eqa
This yields
\bqa
v_0&=&{1\over3}(\mu_B-\mu_S)\mathbb{1}+
{1\over2}\mu_I\lambda_3+{1\over\sqrt{3}}\mu_S\lambda_8\;.
\label{v0}
\eqa
%where $\lambda_i$ ($i=1,2,3...8$) are the Gell-Mann matrices
%which are normalized as $\text{Tr}\lambda_j\lambda_j=2\delta_{ij}$.
We note that the $\mu_B$-dependent term in Eq.~(\ref{v0})
commutes with $\Sigma$ and $\Sigma^{\dagger}$
in Eqs.~(\ref{nabla0})--~(\ref{nablas}) and so
the baryon chemical potential drops completely out of the
chiral Lagrangian. This reflects the fact that we have only included
the mesonic octet, which has zero baryonic charge.
We therefore set $\mu_B=0$ in the remainder of the paper.

\subsection{Next-to-leading order Lagrangian}
In order to perform calculations beyond tree level, we must go to
next-to-leading order in the low-energy expansion and consider
the terms that contribute to ${\cal L}$ at ${\cal O}\left(p^4\right)$.
There are twelve operators in ${\cal L}_4$~\cite{gasser2}, but only eight of
them are relevant for the present calculations.
They are
\bqa\nonumber
{\cal L}_4&=& L_1\left({\rm Tr}
\left[\nabla_{\mu}\Sigma^{\dagger}\nabla^{\mu}\Sigma\right]\right)^2
%\\ \nonumber&&
+L_2{\rm Tr}\left[\nabla_{\mu}\Sigma^{\dagger}\nabla_{\nu}\Sigma\right]
    {\rm Tr}\left[\nabla^{\mu}\Sigma^{\dagger}\nabla^{\nu}\Sigma\right]
\\&& \nonumber
+L_3{\rm Tr}\left[(\nabla_{\mu}\Sigma^{\dagger}\nabla^{\mu}\Sigma)
(\nabla_{\nu}\Sigma^{\dagger}\nabla^{\nu}\Sigma)\right]
+L_4{\rm Tr}\left[\nabla_{\mu}\Sigma^{\dagger}\nabla^{\mu}\Sigma\right]
{\rm Tr}\left[\chi^{\dagger}\Sigma+\chi\Sigma^{\dagger}\right]
\\ &&\nonumber
+L_5{\rm Tr}\left[\left(\nabla_{\mu}\Sigma^{\dagger}\nabla^{\mu}\Sigma\right)
\left(\chi^{\dagger}\Sigma+\chi\Sigma^{\dagger}\right)\right]
+L_6\left({\rm Tr}\left[\chi^{\dagger}\Sigma
    +\chi\Sigma^{\dagger}\right]\right)^2
%+L_7\left[{\rm Tr}\left(\chi^{\dagger}\Sigma
%    -\chi\Sigma^{\dagger}\right)\right]^2
\\ &&
+L_8{\rm Tr}\left[\chi^{\dagger}\Sigma \chi^{\dagger}\Sigma
+  \chi\Sigma^{\dagger}\chi\Sigma^{\dagger}\right]
+H_2{\rm Tr}\left[\chi^{\dagger}\chi\right]\;.
\label{lag}
\eqa
%\end{widetext}
where $L_i$ and $H_i$
are unrenormalized couplings. %~\cite{gasser1}.
The relations between the  bare and renormalized couplings
$L_{i}^r(\Lambda)$ and $H_{i}^r(\Lambda)$
are \bqa
L_{i}&=&L_{i}^r(\Lambda)+
\Gamma_i\lambda\;,
\label{lowl}
\\
  H_i&=&H_{i}^r(\Lambda)
+\Delta_i\lambda\;,
  \label{highl}
  \eqa
  where
  \bqa
  \lambda&=&
  -{\Lambda^{-2\epsilon}\over2(4\pi)^2}\left[{1\over\epsilon}+1\right]\;.
  \eqa
  Here $\Gamma_i$ and $\Delta_i$ are constants and 
  $\Lambda$ is the renormalization scale in the modified minimal
  substraction scheme $\overline{\rm MS}$.
  The renormalized couplings satisfy
  the renormalization group equations
  \bqa
  \label{rgeq0}
\Lambda{d\over d\Lambda}L_i^r&=&-\frac{\Gamma_{i}}{(4\pi)^{2}}\;,
\hspace{1cm}
\Lambda{d\over d\Lambda}H_i^r=-\frac{\Delta_{i}}{(4\pi)^{2}}\;.
\eqa
These are obtained by differentiation of 
Eqs~(\ref{lowl})--(\ref{highl}) 
noting that the bare parameters are independent of the scale $\Lambda$.
The solutions are
\begin{align}
  L_i^r(\Lambda)&=L_i^r(\Lambda_0)-{\Gamma_i\over2(4\pi)^2}
                  \log{\Lambda^2\over\Lambda_0^2}\;,
&H_i^r(\Lambda)=H_i^r(\Lambda_0)-{\Delta_i\over2(4\pi)^2}
                  \log{\Lambda^2\over\Lambda_0^2}\;,
\label{rgeq}
\end{align}
where $\Lambda_0$ is a reference scale.
We note that the contact term $H_2{\rm Tr}[\chi^{\dagger}\chi]$ gives
a constant contribution to the effective potential which is the same in
all phases. We keep it, however, since it is needed to show the scale
independence of the final result for the effective potential.

In three-flavor QCD, the constants $\Gamma_i$ and $\Delta_i$ are
  \begin{align}
&  \Gamma_{1}=\frac{3}{32}\;,& \Gamma_{2}&=\frac{3}{16}\;,& \Gamma_{3}&=0\;,
&  \Gamma_{4}&={1\over8}\;,
  \\
    &  \Gamma_{5}=\frac{3}{8}\;,& \Gamma_{6}&=\frac{11}{144}\;,
%                                                         &  \Gamma_{7}&=0\;,
    &
      \Gamma_{8}&={5\over48}\;,
%    \\
    &    \Delta_2  &={5\over24}    \;.
  \end{align}
%    In writing the NLO Lagrangian above, we have ignored
%    contributions at finite isospin through the Wess-Zumino-Witten (WZW)
%    Lagrangian, which is of the form 
% \bqa
% \mathcal{L}^{\rm ext}_{\rm WZW}&=&
% \frac{-\epsilon^{\mu\nu\alpha\beta}}{48\pi^{2}}
% {\rm Tr}\left [\mu\delta_{\beta0}(L_{\mu}L_{\nu}L_{\alpha}
%   -R_{\mu}R_{\nu}R_{\alpha})
% \right ]\;,
% \eqa
% with
% \bqa
% \mu&=&\frac{\mu_{B}-\mu_{I}}{3}\mathbb{1}
 %{\sqrt{6}}\lambda_{0}
% +\frac{\mu_{I}}{2}\lambda_{3}
%  +\frac{\mu_{I}}{\sqrt{3}}\lambda_{8}\;,\\
%\lambda_{0}&=&\sqrt{\frac{2}{3}}\mathbb{1}\\
%L_{\mu}&=&\Sigma\partial_{\mu}\Sigma^{\dagger}\;,\\
%R_{\mu}&=&\Sigma^{\dagger}\partial_{\mu}\Sigma\;,
%\eqa
%with the leading contribution at $\mathcal{O}(p^{4})$.
%There is also a separate contribution at zero external field at the same order
%but neither of them contribute to the quantities we compute at NLO.

\section{Ground state and fluctuations}
In this section, we will discuss the 
phase structure of the theory
as a function of the chemical potentials $\mu_I$ and $\mu_S$.
We will also discuss how to parametrize the fluctuations above the ground
state. 

The most general $SU(3)$ matrix for the ground state can be written as
\bqa
\Sigma_{\alpha}&=&e^{i\alpha\hat{\phi}_i\lambda_i}\;,
\label{grall}
\eqa
where $\alpha$ is a rotation
angle, $\hat{\phi}_i$ are variational parameters and a sum over the repeated index $i$ is implied.
In order to ensure the normalization of the ground state,
$\Sigma_{\alpha}\Sigma_{\alpha}^{\dagger}=\mathbb{1}$, the
coefficients must satisfy $\sum_{i}\hat{\phi}_i^2=1$.
However, depending on the chemical potentials, we expect that the
ground state takes a certain form, i.e. that it is rotated in a specific way.
For example, in the case $\mu_S=0$, we expect pion condensation
for $|\mu_I|>m_{\pi}$~\cite{son} and that the two-flavor results carry over.
We therefore briefly review the two-flavor case first.
Here the ground state can be written as~\cite{son}
\bqa
\Sigma_{\alpha}&=&e^{i\alpha\hat{\phi}_i\tau_i}
=\cos\alpha+i\hat{\phi}_i\tau_i\sin\alpha\;,
\label{sigma2}
\eqa
where $\tau_i$ are the Pauli matrices and $\hat{\phi}_i$ are again variational
parameters. 
The static part of the ${\cal O}(p^2)$ Hamiltonian ${\cal H}_2$
reads
%\bqa
%{\cal H}_2^{\rm static}&=&
%{f^2\over4}\text{Tr}[v_0,\Sigma_{\alpha},][v_0,\Sigma^{\dagger}_{\alpha}]
%-{f^2\over2}B_0\text{Tr}[M\Sigma_{\alpha}+M\Sigma_{\alpha}^{\dagger}]\;,
%\label{stato}
%\eqa
\begin{equation}
{\cal H}_2^{\rm static}=
{f^2\over4}\text{Tr}[v_0,\Sigma_{\alpha}][v_0,\Sigma^{\dagger}_{\alpha}]
-{f^2\over2}B_0\text{Tr}[M\Sigma_{\alpha}+M\Sigma_{\alpha}^{\dagger}]\;,
\label{stato}
\end{equation}
where in the two-flavor case $v_0={1\over2}\tau_3\mu_I$, cf. Eq.~(\ref{v0})
and $M={\rm diag}(m_u,m_d)={\rm diag}(m,m)$.
The first term in Eq.~(\ref{stato})
can be written as
\bqa%\nonumber
{\cal H}_2^{\rm static\,(a)}&=&{f^2\over4}\text{Tr}
[v_0,\Sigma_{\alpha}][v_0,\Sigma_{\alpha}^{\dagger}]
%={1\over2}f^2\text{Tr}[v_0\Sigma v_0\Sigma^{\dagger}-1]
%\\&&
={f^2\over8}\mu_I^2\text{Tr}[\tau_3\Sigma_{\alpha}
\tau_3\Sigma_{\alpha}^{\dagger}-\mathbb{1}]\;.
\label{firtterm}
\eqa
This form suggests that ${\cal H}_2^{\rm static\,(a)}$ favors
directions that anticommute with $\tau_3$~\cite{son}.
Substituting Eq.~(\ref{sigma2}) into Eq.~(\ref{firtterm}), this expectation
is made explicit,
${\cal H}_2^{\rm static\,(a)}=
-{1\over2}f^2\mu_I^2\sin^2\alpha(\hat{\phi}_1^2+\hat{\phi}_2^2)$.
Evaluating the other term in Eq.~(\ref{stato}), we find
\bqa
{\cal H}_2^{\rm static}&=&-2f^2B_0m\cos\alpha
-{1\over2}f^2\mu_I^2\sin^2\alpha(\hat{\phi}_1^2+\hat{\phi}_2^2)\;.
\label{toth}
\eqa
The first term favors $\alpha=0$, i.e. the vacuum state $\Sigma_0=\mathbb{1}$,
and it is clear that there is competition between the two terms in
Eq.~(\ref{toth}). We notice that the static energy only depends
on $\hat{\phi}_1^2+\hat{\phi}_2^2$, and it is minimized by
setting $\hat{\phi_3}=0$. Without loss of generality and for
later convenience, we can choose $\hat{\phi}_1=1$ and $\hat{\phi}_2=0$.
The rotated vacuum Eq.~(\ref{sigma2}) can then be written as
\bqa
\Sigma_{\alpha}&=&A_{\alpha}\Sigma_0A_{\alpha}\;,
\eqa
where
\bqa
A_{\alpha}=e^{i{\alpha\over2}\tau_1}=\cos\mbox{$\alpha\over2$}
+i\tau_1\sin\mbox{$\alpha\over2$}\;.
\eqa
Minimizing Eq.~(\ref{toth}) with respect to $\alpha$, we find two phases,
$\alpha=0$ for $2B_0m<\mu_I^2$ and
$\cos\alpha={{2B_0m\over\mu_I^2}}$ for $2B_0m>\mu_I^2$. The first phase is the
vacuum phase and the second phase consists of a condensate of charged
pions.
%For $\mu_I>0$ the condensate is of $\pi^+$ and for $\mu_I<0$, it is of $\pi^-$.

In analogy with the two-flavor case, we expect that
pion condensation in the three-flavor case can be
captured by writing Eq.~(\ref{grall}) as~\footnote{
  $\lambda_1$  plays the role of $\tau_1$ and $\lambda_2$ that of
  $\tau_2$. We are free to choose any linear combination of the two and we
  choose
$\lambda_2$.}
\bqa
\Sigma_{\alpha}^{\pi^{\pm}}&=&A_{\alpha}\Sigma_0A_{\alpha}\;,
\eqa
where %$A_{\alpha}=^{i{\alpha\over2}\lambda_2}$.
\bqa
A_{\alpha}&=&e^{i{\alpha\over2}\lambda_2}
={1+2\cos{\alpha\over2}\over3}\mathbb{1}
+i\lambda_2\sin\mbox{$\alpha\over2$}
+{\cos{\alpha\over2}-1\over\sqrt{3}}\lambda_8\;.
\eqa
The rotated ground state can also be conveniently written as
\bqa
\Sigma_{\alpha}^{\pi^{\pm}}&=&
\begin{pmatrix}
\cos\alpha&\sin\alpha&0\\
-\sin\alpha&\cos\alpha&0\\
0&0&1
\end{pmatrix}\;,
\eqa
which shows that the rotation does not affect the $s$-quark.
The symmetry breaking pattern in this case is
\bqa
U(1)_{I_3}\times U(1)_Y\times U(1)_B&\rightarrow&
U(1)_Y\times U(1)_B\;.
\eqa
Since $U(1)_Q\not\subset U(1)_Y\times U(1)_B$, electric charge $Q$ is
also broken and the system is both a superfluid and a superconductor.

We next consider kaon condensation in three-flavor $\chi$PT.
Depending on the values of $\mu_I$ and $\mu_S$, either the charged
kaons or neutral kaons condense. If
$|{1\over2}\mu_I+\mu_S|=|\mu_u-\mu_s|>m_K$, we expect either
$K^+$ or $K^-$
to condense depending on the sign.
If $|-{1\over2}\mu_I+\mu_S|=|\mu_d-\mu_s|>m_K$, we expect $K^0$
or $\bar{K}^0$ to condense depending on the sign.
In the case of charged kaon condensation,
$\lambda_4$ and $\lambda_5$ replace $\lambda_1$ and $\lambda_2$, respectively,
and without loss of generality we can write
%since we consider a condensate  with $u$ and anti-$s$ quarks.
%We then write
$\Sigma_{\alpha}^{K^{\pm}}=e^{i{\alpha\over2}\lambda_5}\Sigma_0e^{i{\alpha\over2}\lambda_5}$.
%$  \Sigma_{\alpha}^{K^{+}}=A_{\alpha}^{K}\Sigma_0A_{\alpha}^{K}$
%where $A_{\alpha}^{K}=e^{i{\alpha\over2}\lambda_5}$.
The rotated ground state takes the form
  \bqa\nonumber
  \Sigma_{\alpha}^{K^\pm}
&=&\nonumber
{1+2\cos\alpha\over3}\mathbb{1}
+{\cos\alpha-1\over2\sqrt{3}}\left(\sqrt{3}\lambda_3-\lambda_8\right)
+i\lambda_5\sin\alpha
\\ &=&
\begin{pmatrix}
\cos\alpha&0&\sin\alpha\\
0&1&0\\
-\sin\alpha
&0&\cos\alpha
\end{pmatrix}\;.
\eqa
The symmetry-breaking pattern is
\bqa
U(1)_{I_3}\times U(1)_Y\times U(1)_B&\rightarrow&
U(1)_Y\times U(1)_B\;.
\eqa
Again, since the $U(1)_Q\not\subset U(1)_Y\times U(1)_B$, electric charge is
spontaneously broken and the superfluid is also a superconductor.

Finally, in the case of neutral kaon condensation 
the rotated ground state is
$\Sigma_{\alpha}^{K^0/\bar{K}^0}=e^{i{\alpha\over2}\lambda_7}\Sigma_0e^{i{\alpha\over2}\lambda_7}$, or
  \bqa\nonumber
  \Sigma_{\alpha}^{K^0/\bar{K}^0}
&=&\nonumber
{1+2\cos\alpha\over3}\mathbb{1}
+{1-\cos\alpha\over2\sqrt{3}}\left(\sqrt{3}\lambda_3+\lambda_8\right)
+i\lambda_7\sin\alpha
\\ &=&
\begin{pmatrix}
1&0&0\\
0&  \cos\alpha&\sin\alpha\\
0&-\sin\alpha
&\cos\alpha
\end{pmatrix}\;.
\eqa
The symmetry-breaking pattern is now
\bqa
U(1)_{I_3}\times U(1)_Y\times U(1)_B&\rightarrow&
U(1)_Q\times U(1)_B\;,
\eqa
implying that the superfluid is not a superconductor.

While we have considered the possibility of a single species condensing, in principle it is possible for the ground state to have simultaneous condensation of multiple mesons. However, explicit calculations in Ref.~\cite{massrev,kogut3} that include the possibility of multiple rotations into multiple condensed phases show that such phases are not the global minima except on the first order transition line. We discuss the line at the end of this section.

We now return to the evaluation of the static Hamiltonian ${\cal H}_2$.
In the case of pion condensation,
the static Hamiltonian reduces to
\bqa
{\cal H}_2&=&-2f^2B_0m\cos\alpha-f^2B_0m_s-{1\over2}f^2\mu_I^2\sin^2\alpha\;.
\eqa
The minimum of the static Hamiltonian is
\begin{align}
\cos\alpha&=1\;,&\mu_I^2<&2B_0m
\\
\cos\alpha&={2B_0m\over\mu_I^2}\;,&\mu_I^2>&2B_0m\;.
\end{align}
The ground-state energy in the vacuum and pion-condensed phase is
\begin{align}
{\cal H}_2&=-f^2B_0(2m+m_s)%=-{1\over2}f^2m_{\pi,0}^2-f^2m_{K,0}^2
\;,&\mu_I^2<&2B_0m\;,\\
%\nonumber
{\cal H}_2&=-{(2fB_0m)^2\over\mu_I^2}-f^2B_0m_s-{1\over2}f^2\mu_I^2\left(
  1-{(2B_0m)^2\over\mu_I^4}\right)
%&&-{f^2\over2\mu_I^2}\left(\mu_I^2-m_{\pi}^2\right)^2
%-f^2(m_K^2-m_{\pi}^2)
\;,&\mu_I^{2}>&2B_0m
\label{firsth}
\;.
\end{align}
%For $\beta={\pi\over2},\gamma=0$,
In the case of charged kaon condensation,
the static Hamiltonian reduces to
\bqa%\nonumber
{\cal H}_2&=&-f^2B_0m(1+\cos\alpha)-f^2B_0m_s\cos\alpha
%\\ &&
-{1\over2}f^2\left(\mbox{$1\over2$}\mu_I+\mu_S\right)^2\sin^2\alpha\;.
\eqa
The minimum of the static Hamiltonian is
\begin{align}
\cos\alpha&=1\;,&\left(\mbox{$1\over2$}\mu_I+\mu_S\right)^2<B_0(m+m_s)
\\
  \cos\alpha&={B_0(m+m_s)\over(\mbox{$1\over2$}\mu_I+\mu_S)^2}
              \;,& 
\left(\mbox{$1\over2$}\mu_I+\mu_S\right)^2>B_0(m+m_s)                           
                            %\mu_S>Gm-\mbox{$1\over2$}\mu_I
                            \;.
\end{align}
The ground-state energy in the vacuum and the charged kaon-condensed phase is
\bqa%\nonumber
{\cal H}_2&=&-f^2B_0(2m+m_s)%=-{1\over2}f^2m_{\pi,0}^2-f^2m_{K,0}^2\;,
\;,(\mbox{$1\over2$}\mu_I+\mu_S)^2<B_0(m+m_s)
\;,\\
\nonumber
{\cal H}_2&=&-f^2B_0m-{f^2B_0^2(m+m_s)^2\over(\mbox{$1\over2$}\mu_I+\mu_S)^2}
%\\ \nonumber &&
-{1\over2}f^2\left(\mbox{$1\over2$}\mu_I+\mu_S\right)^2
\left(1-{B_0^2(m+m_s)^2\over(\mbox{$1\over2$}\mu_I+\mu_S)^4}\right)\;,
%\\% \nonumber&=&
%-{1\over2(\mbox{$1\over2$}\mu_I+\mu_S)^2}
%\left[m_K^2-(\mbox{$1\over2$}\mu_I+\mu_S)^2\right]^2
\\ &&
%-f^2(m_K^2-m_{\pi}^2)
\hspace{0.1cm}(\mbox{$1\over2$}\mu_I+\mu_S)^2>B_0(m+m_s)
\;.
\label{lasth}
\eqa
Finally, we consider the case of % $\beta=\gamma={\pi\over2}$, which corresponds to
condensation of neutral kaons. The results for this phase can be
obtained from the results of the phase of condensed charged kaons by
the substitution $\mu_I\rightarrow-\mu_I$ since
$-{1\over2}\mu_I+\mu_S=\mu_d-\mu_s$.
In order to find the global minimum, we must compare
Eqs.~(\ref{firsth}) and ~(\ref{lasth}) in the region
$|\mu_I|>m_{\pi,0}$ and $|\mbox{$1\over2$}\mu_I+\mu_S|>m_{K,0}$.
The boundary between the pion-condensed phase and the kaon-condensed
phase is then given by equating these expressions.
This yields
\bqa
-{\left(\mu_I^2-m_{\pi,0}^2\right)^2\over2\mu_I^2}
%-f^2(m_K^2-m_{\pi}^2)
&=&-{
  \left[m_K^2-(\mbox{$1\over2$}\mu_I+\mu_S)^2\right]^2
  \over2(\mbox{$1\over2$}\mu_I+\mu_S)^2}\;,
%\\ &&-f^2(m_K^2-m_{\pi}^2)
\eqa
or
\bqa
|\mbox{$1\over2$}\mu_I+\mu_S|
&=&{\mu_I^2-m_{\pi,0}^2+\sqrt{(\mu_I^2-m_{\pi,0}^2)^2+4\mu_I^2m_{K,0}^2}\over2\mu_I}
\label{phaseline}
\;,
\eqa
where we used the tree-level relations $m_{\pi,0}^2=2B_0m$ and $m_{K,0}^2=B_0(m+m_s)$.
We will return to the phase diagram in the $\mu_I$--$\mu_S$ plane in
Sec.~\ref{pd}.

\subsection{Parametrizing Fluctuations}
Since we want to study the thermodynamics of the
pion-condensed and kaon-condensed phases including leading-order
quantum corrections, it is natural to expand the
chiral perturbation theory Lagrangian around the relevant ground state.
The Goldstone manifold as a consequence of chiral symmetry breaking is
$SU(3)_{L}\times SU(3)_{R}/SU(3)_{V}$. We will focus on the
pion-condensed phase for simplicity. The remarks below also apply to
the kaon-condensed phases.
Following Refs.~\cite{kim,us}, we write
\bqa
\Sigma&=L_{\alpha}\Sigma_{\alpha}R_{\alpha}^{\dagger}\;,
\label{sigmas}
\eqa
with 
\bqa
\label{lrot}
L_{\alpha}&=&A_{\alpha}UA^{\dagger}_{\alpha}\;,\\
R_{\alpha}&=&A_{\alpha}^{\dagger}U^{\dagger}A_{\alpha}\;.
\label{rrot}
\eqa
We emphasize that the fluctuations parameterized by $L_{\alpha}$ and $R_{\alpha}$
around the ground state depend on $\alpha$ since the broken generators
(of QCD) need to be rotated appropriately as the condensed vacuum rotates with
the angle $\alpha$~\cite{kim}.
%~\footnote{Consider e.g.
%  a theory with an $SO(3)$ symmetric Lagrangian with the ground state
%  picking up a vev say in the $z$-direction. If the vev is rotated to the
%  $y$-direction, then the (un)broken generators must be rotated
%  accordingly.} 
In the present case,
$U$ is an $SU(3)$ matrix that parameterizes the fluctuations around the vacuum,
\bqa
U=\exp\left (i\frac{\phi_a\lambda_a}{2f} \right )\;.
\eqa
%where $\lambda_a$ ($a=1,2,3...8$) are the Gell-Mann matrices that satisfy
%${\rm Tr}(\lambda_a\lambda_b)=2\delta_{ab}$.
%We also define $\lambda_0=\sqrt{{2\over3}}\mathbb{1}$, where $\mathbb{1}$
%is the unit matrix.
With the parameterizations stated above, we get
\bqa
\label{parametrization}
\Sigma&=&A_{\alpha}(U\Sigma_{0}U)A_{\alpha}\;.
\eqa
This parameterization not only produces the
correct linear terms that vanish when evaluated at the
minimum of the static Hamiltonian $\mathcal{O}(p^{2})$, the
divergences of the one-loop vacuum diagrams also cancel using
counterterms from the $\mathcal{O}(p^{4})$ Lagrangian. Furthermore, the
parametrization produces a Lagrangian that is canonical in the fluctuations and
has the correct limit when $\alpha=0$, whereby 
\bqa
\Sigma&=&U\Sigma_{0}U=U^{2}=\exp\left
  (i\frac{\phi_{a}\lambda_{a}}{f} \right )\; ,
\eqa
as expected. If one expands the Lagrangian using the
parametrization
$\Sigma=L\Sigma_{\alpha}R=U\Sigma_{\alpha}U=UA_{\alpha}\Sigma_0A_{\alpha}U$
instead of Eq.~(\ref{sigmas}), the kinetic terms of the Lagrangian
are non-canonical. By a field redefinition that depends on the
chemical potentials, these terms
can be made canonical. However, calculating the leading corrections
to the tree-level potential, it can be shown that the ultraviolet divergences
can be eliminated by renormalization only at the minimum of the classical
potential.~\footnote{Renormalization of the effective potential
  is carried out by renormalizing the low-energy constants in the NLO
  static Lagrangian, see Sec.~\ref{effpot22}.}
Thus one cannot find the minimum of the next-to-leading order effective
potential as a function of $\alpha$, showing that this parametrization
is erroneous. Let us finally take a look at the rotated generators.
To linear order in the $\phi_i$, 
an infinitesimal fluctuation can be written as
\bqa%\nonumber
L_{\alpha}%&=&A_{\alpha}UA_{\alpha}^{\dagger}
&=&
%\left[\cos{\alpha\over2}+i\lambda_2\sin{\alpha\over2}\right]
\begin{pmatrix}
\cos{\alpha\over2}&\sin{\alpha\over2}&0\\
-\sin{\alpha\over2}&\cos{\alpha\over2}&0\\
0&0&1
\end{pmatrix}
\left[1+i{\phi_i\lambda_i\over2f}\right]
\begin{pmatrix}
\cos{\alpha\over2}&-\sin{\alpha\over2}&0\\
\sin{\alpha\over2}&\cos{\alpha\over2}&0\\
0&0&1
\end{pmatrix}
% \left[\cos{\alpha\over2}-i\lambda_2\sin{\alpha\over2}\right]
\;.
\label{flucti}
\eqa
Using the (anti)commutator relations of the Gell-Mann matrices,
Eq.~(\ref{flucti}) takes the form
\bqa\nonumber
L_{\alpha}&=&1+
{i\phi_1\over2f}(\cos\alpha\lambda_1+\sin\alpha\lambda_3)
+{i\phi_2\lambda_2\over2f}
+{i\phi_3\over2f}(\cos\alpha\lambda_3-\sin\alpha\lambda_1)
\\ && \nonumber 
+{i\phi_4\over2f}
\left(\cos\mbox{$\alpha\over2$}\lambda_4-\sin\mbox{$\alpha\over2$}
  \lambda_6\right)
+{i\phi_5\over2f}\left(\cos\mbox{$\alpha\over2$}\lambda_5
  -\sin\mbox{$\alpha\over2$}
  \lambda_7\right)
+{i\phi_6\over2f}\left(\cos\mbox{$\alpha\over2$}\lambda_6
  +\sin\mbox{$\alpha\over2$}
  \lambda_4\right)
\\ &&
+{i\phi_7\over2f}\left(\cos\mbox{$\alpha\over2$}\lambda_7+
  \sin\mbox{$\alpha\over2$}
  \lambda_5\right)
+{i\phi_8\lambda_8\over2f}\;.
\eqa
The linear combinations
$\lambda_1^{\prime}=(\cos\alpha\lambda_1+\sin\alpha\lambda_3)$,
$\lambda_2^{\prime}=\lambda_2$, etc
can be thought of as rotated generators,
some of them, however, only by half the angle.
The rotated generators $\lambda_i^{\prime}$
satisfy the same (anti)commutation relations as do $\lambda_i$
To all orders in $\alpha$, we then have
\bqa
L_{\alpha}&=&\exp\left({i\phi_i\lambda_i^{\prime}\over2f}\right)\;.
\eqa

\subsection{Leading-order Lagrangian}
Using the parameterization Eq.~(\ref{parametrization}) discussed above, we
can write down the Lagrangian in terms of the fields $\phi_{a}$, which
parametrizes the Goldstone manifold. The leading-order terms in the
low-energy expansion are given by ${\cal L}_2$, which can be
expanded as a power series in the fields
\bqa
\mathcal{L}_{2}&=&\mathcal{L}_2^{\rm linear}%^{\rm cond}
+\mathcal{L}_2^{\rm static}%^{\rm cond}
+\mathcal{L}_2^{\rm quadratic}%^{\rm cond}
+\cdots\;
\label{lolag}
\eqa
where the ellipses indicate terms that are cubic or higher order
in the fields. We will carry out the expansion for the normal phase,
the pion-condensed phase, and the 
charged kaon-condensed
phase. Similar results can be obtained for the neutral kaon-condensed
phase.

\subsubsection{Normal Phase}
In the normal phase, 
the different terms in Eq.~(\ref{lolag}) are
%\begin{widetext}
\bqa
\label{nen}
\mathcal{L}_{2}^{\rm static}&=&f^2B_0(2m+m_s)%+\frac{1}{2}f^{2}\mu_{I}^{2}
\;,
\\
  \mathcal{L}_{2}^{\rm linear}&=&0\;,\\ \nonumber
  \mathcal{L}_{2}^{\rm quadratic}&=&\frac{1}{2}\partial_{\mu}\phi_{a}
  \partial^{\mu}\phi^a
  -\frac{1}{2}\left(2B_0m-\mu_{I}^{2}\right)
    \left(\phi_{1}^{2}+\phi_{2}^{2}\right)
  -\frac{1}{2}(2B_0m)\phi_{3}^{2}\\ \nonumber
&&  -\frac{1}{2}\left [B_{0}(m+m_{s})-\left(\frac{1}{2}\mu_{I}+\mu_{S} \right )^{2}\right ](\phi_{4}^{2}+\phi_{5}^{2})\\
\nonumber
&&-\frac{1}{2}\left [B_{0}(m+m_{s})-\left(-\frac{1}{2}\mu_{I}+\mu_{S} \right )^{2}\right ](\phi_{6}^{2}+\phi_{7}^{2})
    \\ \nonumber    &&
    -\frac{B_0(m+2m_{s})}{3}\phi_{8}^{2}
    +\mu_{I}(\phi_{1}\partial_{0}\phi_{2}-\phi_{2}\partial_{0}\phi_{1})+
\left (\mbox{$1\over2$}
\mu_I+\mu_{S} \right )\left (\phi_{4}\partial_{0}
  \phi_{5}-\phi_{5}\partial_{0}\phi_{4}\right )
\\ &&
+
\left (-\mbox{$1\over2$}\mu_I+\mu_{S} \right )\left (\phi_{6}\partial_{0}
    \phi_{7}-\phi_{7}\partial_{0}\phi_{6}\right )\ .
  \eqa
%\end{widetext}
The inverse propagator is block diagonal and can be written as
\bqa
  D^{-1}&=&
\begin{pmatrix}
D^{-1}_{12}&0&0&0&0&\\
0&P^2-m_3^2&0&0&0&\\
0&0&D^{-1}_{45}&0&0&\\
0&0&0&D^{-1}_{67}&0&\\
0&0&0&0&P^{2}-m_{8}^{2}\\
\end{pmatrix}\;,\\
      m_3^2&=&2B_0m\;,\\
  m_8^2&=&{2B_0(m+2m_s)\over3}\;,
\eqa
where $P=(p_0,p)$ is the four-momentum and $P^2=p_0^2-p^2$. 
The submatrices are
\begin{align}
D^{-1}_{12}&=
\begin{pmatrix}
P^{2}-m_{1}^{2}&ip_{0}m_{12}\\
-ip_{0}m_{12}&P^{2}-m_{2}^{2}\\
\end{pmatrix}\;,
&
  D^{-1}_{45}&=
\begin{pmatrix}
P^{2}-m_{4}^{2}&ip_{0}m_{45}\\
-ip_{0}m_{45}&P^{2}-m_{5}^{2}\\
\end{pmatrix}\;,
\\
    D^{-1}_{67}&=
\begin{pmatrix}
P^{2}-m_{6}^{2}&ip_{0}m_{67}\\
-ip_{0}m_{67}&P^{2}-m_{7}^{2}\\
\end{pmatrix}\;,
\end{align}
The masses are
\bqa
m_1^2&=&2B_0m-\mu_I^2%=m_{\pi,0}^2-\mu_I^2
\;,\\
  m_2^2&=&m_1^2\;,\\
  m_{12}&=&2\mu_I\;,
\\
m_{4}^2&=&B_0(m+m_s)-\left(\mbox{$1\over2$}\mu_I+\mu_S\right)^2
%=m_{K,0}^2-\left(\mbox{$1\over2$}\mu_I+\mu_S\right)^2
\;,\\
m_5^2&=&m_4^2\;,\\
m_{45}&=&\mu_I+2\mu_S\;,
     \\
     m_{6}^{2}&=&B_{0}(m+m_{s})-\left(-\frac{1}{2}\mu_{I}+\mu_{S} \right )^{2}\;,\\
%     G(m+m_s)-\left(\mbox{$1\over2$}\mu_I+\mu_S\right)^2
  m_7^2&=&m_6^2\;,\\
  m_{67}&=&-\mu_I+2\mu_S\;.
  \eqa
The dispersion relations for the charges mesons are
\bqa
\label{nspec1}
E_{\pi^{\pm}}&=&\sqrt{p^2+2B_0m}\mp\mu_I
=\sqrt{p^2+m_{\pi,0}^2}\mp\mu_I
\;,\\
E_{\pi^0}&=&\sqrt{p^2+2B_0m}=\sqrt{p^2+m_{\pi,0}^2}\;,\\
E_{K^{\pm}}&=&\sqrt{p^2+B_0(m+m_s)}\mp(\mbox{$1\over2$}\mu_I+\mu_S)
=\sqrt{p^2+m_{K,0}^2}\mp(\mbox{$1\over2$}\mu_I+\mu_S)
\;,\\
E_{K^{0},\bar{K}^{0}}&=&\sqrt{p^2+B_0(m+m_s)}\mp(-\mbox{$1\over2$}\mu_I+\mu_S)
=\sqrt{p^2+m_{K,0}^2}\mp(-\mbox{$1\over2$}\mu_I+\mu_S)\;,\\
E_{\eta}&=&\sqrt{p^2+{2\over3}B_0(m+2m_s)}
=\sqrt{p^2+m_{\eta,0}^2}
\;.
\label{nspec2}
\eqa
The tree-level masses of the pions, kaons, and the $\eta$
are then given by $m^2_{\pi,0}=2B_0m$, $m_{K,0}^2=B_0(m+m_s)$, and
$m_{\eta,0}^2=\mbox{$2\over3$}B_0(m+2m_s)$.
  
\subsubsection{Pion-condensed phase}
In the pion-condensed phase, the different terms in Eq.~(\ref{lolag}) are
%\begin{widetext}
\bqa
\mathcal{L}_{2}^{\rm static}&=&f^2B_0(2m\cos\alpha+m_s)
+\frac{1}{2}f^{2}\mu_{I}^{2}
\sin^{2}\alpha
\label{statlag1}
\\
  \mathcal{L}_{2}^{\rm linear}&=&
\label{linear}
  f(-2B_0m+\mu_{I}^{2}\cos\alpha)\sin\alpha
  \phi_{2}-f\mu_{I}\sin\alpha\partial_{0}\phi_{1}\\ \nonumber
  \mathcal{L}_{2}^{\rm quadratic}&=&\frac{1}{2}\partial_{\mu}\phi_{a}
  \partial^{\mu}\phi_{a}
  -\frac{1}{2}\left (2B_0m\cos\alpha-\mu_{I}^{2}\cos^{2}\alpha
  \right )\phi_{1}^{2}
  \\  && 
  \label{quad}
  \nonumber
  -\frac{1}{2}\left (2B_0m\cos\alpha-\mu_{I}^{2}
    \cos2\alpha \right )\phi_{2}^{2}
  -\frac{1}{2}\left (2B_0m\cos\alpha
    +\mu_{I}^{2}\sin^{2}\alpha \right )\phi_{3}^{2}
\\ && \nonumber
  -\frac{1}{2}\left [B_0(m\cos\alpha+m_{s})
    -{1\over4}\mu_{I}^{2}\cos2\alpha
    -\mu_{I}\mu_{S}\cos\alpha-\mu_{S}^{2}
  \right ](\phi_{4}^{2}+\phi_{5}^{2})\\
  &&\nonumber
  -\frac{1}{2}\left [B_0(m\cos\alpha+m_{s})
    -{1\over4}\mu_{I}^{2}\cos2\alpha
    +\mu_{I}\mu_{S}\cos\alpha-\mu_{S}^{2}
  \right ](\phi_{6}^{2}+\phi_{7}^{2})
  \\  &&\nonumber
  -\frac{B_0(m\cos\alpha+2m_{s})}{3}\phi_{8}^{2}
  +\mu_{I}\cos\alpha(\phi_{1}\partial_{0}\phi_{2}-\phi_{2}\partial_{0}\phi_{1})
\\ && \nonumber
+\left (\mbox{$1\over2$}
  \mu_I\cos\alpha+\mu_{S} \right )\left (\phi_{4}\partial_{0}
\phi_{5}-\phi_{5}\partial_{0}\phi_{4}\right )
+\left (-\mbox{$1\over2$}
  \mu_I\cos\alpha+\mu_{S} \right )\left (\phi_{6}\partial_{0}
  \phi_{7}-\phi_{7}\partial_{0}\phi_{6}\right)\;.
\\ &&
\eqa
%\end{widetext}
% \twocolumngrid
\noindent
%We notice that the linear term vanishes
%at the maximum of ${\cal L}_{2}^{\rm static}$, i.e. at the minimum of the
%tree-level potential, as required.
We get for the inverse
propagator:
%\begin{equation}
%\begin{split}
\bqa
D^{-1}&=&
\begin{pmatrix}
D^{-1}_{12}&0&0&0&0\\
0&p^{2}-m_{3}^{2}&0&0&0\\
0&0&D^{-1}_{45}&0&0\\
0&0&0&D^{-1}_{67}&0\\
0&0&0&0&P^{2}-m_{8}^{2}\\
\end{pmatrix}\;,\\
m_{3}^{2}&=&2B_0m\cos\alpha+\mu_{I}^{2}\sin^{2}\alpha\;, \\
m_{8}^{2}&=&\frac{2B_0(m\cos\alpha+2m_{s})}{3}\;.
\label{invDp}
\eqa
%\end{split}
%\end{equation}
The three different $2\times2$ matrices are given by
\begin{align}
D^{-1}_{12}&=
\begin{pmatrix}
P^{2}-m_{1}^{2}&ip_{0}m_{12}\\
-ip_{0}m_{12}&P^{2}-m_{2}^{2}\\
\end{pmatrix}\;,
  &
D^{-1}_{45}&=
\begin{pmatrix}
P^{2}-m_{4}^{2}&ip_{0}m_{45}\\
-ip_{0}m_{45}&P^{2}-m_{5}^{2}\\
\end{pmatrix}\;,\\
D^{-1}_{67}&=
\begin{pmatrix}
P^{2}-m_{6}^{2}&ip_{0}m_{67}\\
-ip_{0}m_{67}&P^{2}-m_{7}^{2}\\
\end{pmatrix}\;,
\end{align}
where the masses are
\bqa
m_{1}^{2}&=&2B_0m\cos\alpha-\mu_{I}^{2}\cos^{2}\alpha
%=m_{\pi,0}^2\cos\alpha-\mu_{I}^{2}\cos^{2}\alpha
\;, \\
m_{2}^{2}&=&2B_0m\cos\alpha-\mu_{I}^{2}\cos2\alpha
%=m_{\pi,0}^2\cos\alpha-\mu_{I}^{2}\cos2\alpha
\;, \\
m_{12}&=&2\mu_{I}\cos\alpha\;,\\ %\nonumber
m_{4}^{2}&=&B_0(m\cos\alpha+m_{s})
-\frac{\mu_{I}^{2}}{4}\cos2\alpha-\mu_{I}\mu_{S}\cos\alpha-\mu_{S}^{2}
%\\ &=&{1\over2}m_{\pi,0}^2(\cos\alpha-1)+m_{K,0}^2
%-\frac{\mu_{I}^{2}}{4}\cos2\alpha-\mu_{I}\mu_{S}\cos\alpha-\mu_{S}^{2}
\;,\\
m_5^2&=&m_4^2\;,\\
m_{45}&=&\mu_{I}\cos\alpha+2\mu_{S}\;, \\ %\nonumber
m_{6}^{2}&=&B_0(m\cos\alpha+m_{s})
-\frac{\mu_{I}^{2}}{4}\cos2\alpha+\mu_{I}\mu_{S}\cos\alpha-
\mu_{S}^{2}\;,\\
%&=&{1\over2}m_{\pi,0}^2(\cos\alpha-1)+m_{K,0}^2
%-\frac{\mu_{I}^{2}}{4}\cos2\alpha+\mu_{I}\mu_{S}\cos\alpha-\mu_{S}^{2}\;,\\
m_7^2&=&m_6^2\;,\\
m_{67}&=&-\mu_{I}\cos\alpha+2\mu_{S}\;.
\eqa

The quasiparticle dispersion relations can be easily found and read
%\begin{widetext}
  \bqa
  \label{kvasi1}
E_{\pi^0}&=&p^2+m_3^2
\;,\\
\nonumber
E_{\pi^{\pm}}^{2}&=&p^2+{1\over2}\left(m_{1}^{2}+m_{2}^{2}+m_{12}^{2}\right)\\
&\mp&{1\over2}\sqrt{4p^{2}m_{12}^{2}+(m_{1}^{2}+m_{2}^{2}
  +m_{12}^{2})^2-4m_{1}^{2}m_{2}^{2}}\;,\\
  \nonumber
E_{K^{\pm}}^{2}&=&p^2+{1\over2}\left(m_{4}^{2}+m_{5}^{2}+m_{45}^{2}\right)\\
&\mp&{1\over2}\sqrt{4p^{2}m_{45}^{2}+(m_{4}^{2}+m_{5}^{2}
  +m_{45}^{2})^2-4m_{4}^{2}m_{5}^{2}}\;,\\
  \nonumber
E_{K^0,\bar{K}_{0}}^{2}&=&p^2+{1\over2}\left(m_{6}^{2}+m_{7}^{2}+m_{67}^{2}\right)\\
&\mp&{1\over2}\sqrt{4p^{2}m_{67}^{2}+(m_{6}^{2}+m_{7}^{2}
  +m_{67}^{2})^2-4m_{6}^{2}m_{7}^{2}}\;,\\
E_{\eta^0}^2&=&p^2+m_8^2\;.
  \label{kvasi2}
\eqa

\subsubsection{Charged kaon-condensed phase}
In the kaon-condensed phase, the different terms in Eq.~(\ref{lolag}) are
\bqa 
\label{statlag111}
\mathcal{L}_{2}^{\rm static}&=&
f^2B_0[m+(m+m_{s})\cos\alpha]
+\frac{1}{2}f^{2}\left(\mbox{$1\over2$}\mu_I+\mu_S\right)^2
\sin^{2}\alpha\\
\mathcal{L}_{2}^{\rm linear}&=&f\left[-B_{0}(m+m_{s})+
  \left (\tfrac{1}{2}\mu_{I}+\mu_{S} \right )^2\cos\alpha \right ]
\sin\alpha\phi_{5}-f\left (\tfrac{1}{2}\mu_{I}+\mu_{S} \right )\sin\alpha
\partial_{0}\phi_{4}
\label{linear1}
\\ \nonumber
\mathcal{L}_{2}^{\rm quadratic}&=&\frac{1}{2}\partial_{\mu}\phi_{a}
\partial^{\mu}\phi_{a}-\frac{1}{2}\left \{\frac{1}{2}B_{0}\left[3m-m_{s}
    +(m+m_{s})\cos\alpha \right ] \right.\\
\label{quad1}
\nonumber
&&\left.-\frac{1}{16}\left [3\mu_{I}-2\mu_{S}+2\left (\tfrac{1}{2}\mu_{I}
      +\mu_{S} \right )\cos\alpha \right ]^{2}+\frac{1}{4}
  \left (\tfrac{1}{2}\mu_{I}+\mu_{S} \right )^{2}\sin^{2}\alpha\right \}
(\phi_{1}^{2}+\phi_{2}^{2})\\
\nonumber
&&-\frac{1}{2}\left \{\frac{1}{2}B_{0}\left [3m-m_{s}+(m+m_{s})\cos\alpha
  \right ]+\frac{1}{4}\left (\tfrac{1}{2}\mu_{I}+\mu_{S} \right )^{2}\sin^{2}
  \alpha \right\}\phi_{3}^{2}\\
\nonumber
&&-\frac{1}{2}\left\{B_{0}(m+m_{s})\cos\alpha-\left (\tfrac{1}{2}\mu_{I}
    +\mu_{s} \right )^{2}\cos^{2}\alpha \right \}\phi_{4}^{2}\\
\nonumber
&&-\frac{1}{2}\left\{B_{0}(m+m_{s})\cos\alpha-\left (\tfrac{1}{2}\mu_{I}+\mu_{s}
  \right )^{2}\cos2\alpha \right \}\phi_{5}^{2}\\
\nonumber
&&-\frac{1}{2}\left\{\frac{1}{2}B_{0}(m+m_{s})(1+\cos\alpha)-\tfrac{1}{16}
  \left[-3\mu_{I}+2\mu_{S}+2(\tfrac{1}{2}\mu_I+\mu_{S})\cos\alpha \right]^{2}
  \right.\\
  \nonumber
  &&+\left.\tfrac{1}{4}(\tfrac{1}{2}\mu_{I}+\mu_{S})^{2}\sin^{2}\alpha\right\}
  (\phi_{6}^{2}+\phi_{7}^{2})\\
  \nonumber
  &&-\frac{1}{2}\left \{\left[\frac{1}{6}B_{0}(-m+3m_{s}+5(m+m_{s})\cos\alpha)+
      \tfrac{3}{4}\left (\tfrac{1}{2}\mu_{I}+\mu_{S} \right )^{2}\sin^{2}\alpha
    \right] \right \}\phi_{8}^{2}\\
  \nonumber
  &&-\left\{{1\over2\sqrt{3}}B_0(m+m_s)(\cos\alpha-1)+
  {\sqrt{3}\over4}\left(\tfrac{1}{2}\mu_{I}+\mu_{S} \right )^{2}
\sin^{2}\alpha\right\}\phi_3\phi_8\\  \nonumber
  &&+\frac{1}{4}\left [3\mu_{I}-2\mu_{S}+2(\tfrac{1}{2}\mu_{I}+\mu_{S})
    \cos\alpha
 \right ](\phi_{1}\partial_{0}\phi_{2}-\phi_{2}\partial_{0}\phi_{1})\\
\nonumber
 &&+\left (\tfrac{1}{2}\mu_{I}+\mu_{S} \right )\cos\alpha(\phi_{4}\partial_{0}
 \phi_{5}-\phi_{5}\partial_{0}\phi_{4})\\
 &&+\frac{1}{4}\left [-3\mu_{I}+2\mu_{S}+2(\tfrac{1}{2}\mu_{I}+\mu_{S})\cos
   \alpha
\right ](\phi_{6}\partial_{0}\phi_{7}-\phi_{7}\partial_{0}\phi_{6})\;.
\eqa
The inverse propagator is block diagonal and can be written as
  \bqa
  D^{-1}&=
\begin{pmatrix}
D^{-1}_{12}&0&0&0&\\
0&D^{-1}_{38}&0&0&\\
0&0&D^{-1}_{45}&0&\\
0&0&0&D^{-1}_{67}&\\
%0&0&0&0&p^{2}-m_{8}^{2}\\
\end{pmatrix}\;,\\
\label{invDk}
\eqa
where the submatrices are
%\begin{widetext}
  \begin{align}
D^{-1}_{12}&=
\begin{pmatrix}
P^{2}-m_{1}^{2}&ip_{0}m_{12}\\
-ip_{0}m_{12}&P^{2}-m_{2}^{2}\\
\end{pmatrix}\;,
&
D^{-1}_{38}&=
\begin{pmatrix}
P^{2}-m_{3}^{2}&-m_{38}^2\\
-m_{38}^2&P^{2}-m_{8}^{2}\\
\end{pmatrix}\;,
           &
    \\
    \nonumber
    \\            
    D^{-1}_{45}&=
\begin{pmatrix}
P^{2}-m_{4}^{2}&ip_{0}m_{45}\\
-ip_{0}m_{45}&P^{2}-m_{5}^{2}\\
\end{pmatrix}\;,
           &
    D^{-1}_{67}&=
\begin{pmatrix}
P^{2}-m_{6}^{2}&ip_{0}m_{67}\\
-ip_{0}m_{67}&P^{2}-m_{7}^{2}\\
\end{pmatrix}\;.             
  \end{align}
The masses are
\bqa \nonumber
m_{1}^{2}&=&\left \{\frac{1}{2}B_{0}\left[3m-m_{s}+(m+m_{s})\cos\alpha \right ]
\right.\\
&&\left.-\frac{1}{16}\left [3\mu_{I}-2\mu_{S}+2\left (\tfrac{1}{2}\mu_{I}
      +\mu_{S} \right )\cos\alpha \right ]^{2}+\frac{1}{4}\left (\tfrac{1}{2}
    \mu_{I}+\mu_{S} \right )^{2}\sin^{2}\alpha\right \}\;,\\
m_{2}^{2}&=&m_{1}^{2}\;,\\
m_{12}&=&-\frac{1}{2}\left [3\mu_{I}-2\mu_{S}+2(\tfrac{1}{2}\mu_{I}+\mu_{S})
  \cos\alpha \right ]\;,\\
m_{3}^{2}&=&\left \{\frac{1}{2}B_{0}\left [3m-m_{s}+(m+m_{s})\cos\alpha \right ]
  +\frac{1}{4}\left (\tfrac{1}{2}\mu_{I}+\mu_{S} \right )^{2}\sin^{2}\alpha
\right\}\;,\\
m_{4}^{2}&=&\left\{B_{0}(m+m_{s})\cos\alpha-\left (\tfrac{1}{2}\mu_{I}+\mu_{S}
  \right )^{2}\cos^{2}\alpha \right \}\;,\\
m_{5}^{2}&=&\left\{B_{0}(m+m_{s})\cos\alpha-\left (\tfrac{1}{2}\mu_{I}+\mu_{S}
  \right )^{2}\cos2\alpha \right \}\;,\\
m_{45}&=&-2\left (\tfrac{1}{2}\mu_{I}+\mu_{S} \right )\cos\alpha\;,\\
\nonumber
m_{6}^{2}&=&\left\{\frac{1}{2}B_{0}(m+m_{s})(1+\cos\alpha)-\tfrac{1}{16}\left
    [-3\mu_{I}+2\mu_{S}+2(\tfrac{\mu_{I}}{2}+\mu_{S})\cos\alpha \right]^{2}
\right.\\
&&+\left.\tfrac{1}{4}(\tfrac{1}{2}\mu_{I}+\mu_{S})^{2}\sin^{2}\alpha\right\}
\;,\\
m_{7}^{2}&=&m_{6}^{2}\;,\\
m_{67}&=&-\frac{1}{2}\left [-3\mu_{I}+2\mu_{S}+2(\tfrac{1}{2}\mu_{I}+\mu_{S})
  \cos\alpha \right ]\;,\\
m_{8}^{2}&=&\left[\frac{1}{6}B_{0}(-m+3m_{s}+5(m+m_{s})\cos\alpha)+
    \tfrac{3}{4}\left (\tfrac{1}{2}\mu_{I}+\mu_{S} \right )^{2}\sin^{2}\alpha
  \right]\;,\\
  m_{38}^{2}&=&{1\over{{2}\sqrt{3}}}B_0(m+m_s)(\cos\alpha-1)
+{\sqrt{3}\over4}\left(\mbox{$1\over2$}\mu_I+\mu_S\right)^2\sin^2\alpha\;.
\eqa
The quasiparticle dispersion relations can be easily found and read
  \bqa
  \label{kvasi3}
  E_{\pi^0}^2&=&p^2+{1\over2}(m_3^2+m_8^2)+{1\over2}
  \sqrt{(m_3^2-m_8^2)^2+4m_{38}^4}
\;,\\
E_{\pi^{\pm}}^{2}&=&p^2+{1\over2}\left(m_{1}^{2}+m_{2}^{2}+m_{12}^{2}\right)\\
&
\mp&{1\over2}\sqrt{4p^{2}m_{12}^{2}+(m_{1}^{2}+m_{2}^{2}
  +m_{12}^{2})^2-4m_{1}^{2}m_{2}^{2}}\;,\\
E_{K^{\pm}}^{2}&=&p^2+{1\over2}\left(m_{4}^{2}+m_{5}^{2}+m_{45}^{2}\right)\\
&\mp&{1\over2}\sqrt{4p^{2}m_{45}^{2}+(m_{4}^{2}+m_{5}^{2}
  +m_{45}^{2})^2-4m_{4}^{2}m_{5}^{2}}\;,\\
E_{K^0,\bar{K}^0}^{2}&=&p^2+{1\over2}\left(m_{6}^{2}+m_{7}^{2}+m_{67}^{2}\right)\\
&\mp&{1\over2}\sqrt{4p^{2}m_{67}^{2}+(m_{6}^{2}+m_{7}^{2}
  +m_{67}^{2})^2-4m_{6}^{2}m_{7}^{2}}\;,\\
E_{\eta^0}^2&=&p^2+{1\over2}(m_3^2+m_8^2)-{1\over2}
  \sqrt{(m_3^2-m_8^2)^2+4m_{38}^4}\;.
  \label{kvasi4}
\eqa  
The linear terms in the condensed phases are given by Eqs.~(\ref{linear})
and~(\ref{linear1}).
By differentiation with respect to $\alpha$,
it is straightforward to see that the terms vanish at the extremum
of the corresponding static Lagrangian.

\section{Next-to-leading order effective potential}
\label{effpot22}
In this section, we calculate the NLO effective potential in
the three different phases we consider.
At  $\mathcal{O}(p^2)$, the contribution to the effective potential in each
phase is given by evaluating $-{\cal L}_2^{\rm static}$ using
$\Sigma_{\alpha}^{\pi^{\pm}}$, $\Sigma_{\alpha}^{K^{\pm}}$, or
$\Sigma_{\alpha}^{K^{0}/\bar{K}^0}$.
At $\mathcal{O}(p^4)$, there are two contributions to the effective potential.
The first is the Gaussian fluctuations about the ground state, i.e.
the standard one-loop contribution. The second is given by 
evaluating $-{\cal L}_4^{\rm static}$, again using 
$\Sigma_{\alpha}^{\pi^{\pm}}$, $\Sigma_{\alpha}^{K^{\pm}}$, or
$\Sigma_{\alpha}^{K^{0}/\bar{K}^0}$.
The one-loop contribution is
ultraviolet divergent and needs regularization. We regularize the ultraviolet
divergences using dimensional regularization in $d=3-2\epsilon$ dimensions.
The divergences are cancelled by renormalizing the coupling constants
that multiply the operators in ${\cal L}_4$. The sum of the three contributions
is the complete effective potential to $\mathcal{O}(p^4)$ in $\chi$PT.

After going to Euclidean space,
the one-loop contribution to the effective potential of a free massive
boson is given by
\bqa
V_1&=&
{1\over2}\int_P\log\left[P^2+m^2\right]=
{1\over2}\int{dp_0\over2\pi}\int_p\log\left[p_0^2+p^2+m^2\right]\;,
\eqa
where $m$ is the mass and 
the second integral is defined %in dimensional regularization
in $d=3-2\epsilon$ dimensions as
\bqa
\int_{p}&=&\left (\frac{e^{\gamma_{E}}\Lambda^{2}}{4\pi} \right )^{\epsilon}
\int
\frac{d^{d}p}{(2\pi)^{d}}\;,
\label{sumint}
\eqa
and where $\Lambda$ is the renormalization scale associated with the
modified minimal subtraction scheme ($\overline{\rm MS}$).
%and and the integral is in Euclidean space in $d=4-2\epsilon$
%dimension.
Integrating over $P_0$, one finds
\bqa
\label{int1}
V_1&=&{1\over2}\int_p\sqrt{p^2+m^2}
=-{m^4\over4(4\pi)^2}
\left({{\Lambda^2\over m^2}}\right)^{\epsilon}
\left[{1\over\epsilon}+{3\over2}+{\cal O}(\epsilon)\right]\;.
\eqa

\subsection{Normal phase}
The leading-order contribution to the effective potential is minus
the static Lagrangian given in Eq.~(\ref{nen})
%evaluated at $\alpha=0$,
\bqa
V_0&=&-f^2B_0(2m+m_s)
%=-f^2\left({1\over2}m_{\pi,0}^2+m_{K,0}^2\right)
\;.
\eqa
The one-loop contribution to the effective potential is 
\bqa
V_1&=&
{1\over2}\int_p\left[
  E_{\pi^+}+E_{\pi^-}+E_{\pi^0}
+  E_{K^+}+E_{K^-}+E_{K^0}+E_{\bar{K}^0}+E_{\eta^0}\right]\;,
\label{loopie}
\eqa
where the particle energies are given by Eqs.~(\ref{nspec1})--(\ref{nspec2}).
Using Eq.~(\ref{int1}), we can write Eq.~(\ref{loopie}) as
\bqa\nonumber
V_1&=&
-{3\over4(4\pi)^{2}}\left [{1\over\epsilon}+\frac{3}{2} +
  \log\left (\frac{\Lambda^{2}}{{m_{\pi,0}^2}}
  \right ) \right ]\left[2B_0m\right]^2
\\  \nonumber
&&-{1\over(4\pi)^{2}}\left [\frac{1}{\epsilon}+\frac{3}{2} +
  \log\left (\frac{\Lambda^{2}}{{m_{K,0}^2}}
  \right ) \right ]\left[B_0(m+m_s)\right]^2
\\ && 
-{1\over4(4\pi)^{2}}\left [\frac{1}{\epsilon}+\frac{3}{2} +
  \log\left (\frac{\Lambda^{2}}{{m_{\eta,0}^2}}
  \right ) \right ]\left[{2B_0(m+2m_s)\over3}\right]^2\;.
\label{eftnormal}
\eqa
The $\mathcal{O}(p^4)$ contribution from minus the static 
Lagrangian ${\cal L}_4^{\rm static}$ is given by %evaluated at $\alpha=0$ is
\bqa%\nonumber
V_1^{\rm static}&=&
-16L_6B_0^2(2m+m_s)^2-8L_8B_0^2(2m^2+m_s^2)
-4H_2B_0^2(2m^2+m_s^2)
%\\ &=& \nonumber
%-m_{\pi,0}^4\left[6L_6-5L_8-{5\over2}H_2\right]
%-m_{K,0}^4\left[48L_6-8L_8-4H_2\right]
%\\&&+m_{\eta,0}^4\left[18L_6-9L_8-{9\over2}H_2\right]\;.
\label{statcount1}
\eqa
After renormalization, the effective potential is
\bqa\nonumber
V_{\rm eff}&=&-f^2B_0(2m+m_s)
-16L_6^rB_0^2(2m+m_s)^2-8L_8^rB_0^2(2m^2+m_s^2)
-4H_2^rB_0^2(2m^2+m_s^2)
% -{1\over2}f^2\mu_I^2\sin^2\alpha
\\ && \nonumber
%-\left[4{L}_1^r+4{L}_2^r+2{L}_3^r
%  +{1\over16(4\pi)^2}\left({9\over2}+8\log{\Lambda^2\over m_3^2}
%  +\log{\Lambda^2\over\tilde{m}_3^2}\right)
%\right]\mu_I^4\sin^4\alpha
%\\ && \nonumber
%-\left[8{L}_4
%  +{1\over2(4\pi)^2}\left({1\over2}+\log{\Lambda^2\over\tilde{m}_3^2}\right)
%\right](2Gm\cos\alpha+m_s)\mu_I^2\alpha\sin^2\alpha
%\\ && \nonumber
%-\left[8{L}_5
%  +{1\over2(4\pi)^2}\left({3\over2}+4\log{\Lambda^2\over m_3^2}
%  -\log{\Lambda^2\over\tilde{m}_3^2}\right)
%\right]Gm\mu_I^2\cos\alpha\sin^2\alpha
%+G^2m^2\sin^2\alpha\left[16L_8^r-8H_2^r\right]
%\\ && \nonumber
-\left[
  {1\over(4\pi)^2}\left({37\over18}
    +3\log{\Lambda^2\over m_{\pi,0}^2}+
    \log{\Lambda^2\over m_{K,0}^2}
+{1\over9}\log{\Lambda^2\over m_{\eta,0}^2}
\right)\right]B_0^2m^2
\\ && \nonumber
-\left[
{1\over(4\pi)^2}\left({11\over9}    +2\log{\Lambda^2\over m_{K,0}^2}
  +{4\over9}\log{\Lambda^2\over m_{\eta,0}^2}\right)\right]B_0^2mm_s
\\ &&%\nonumber
-\left[
  {1\over(4\pi)^2}\left({13\over18}
    +\log{\Lambda^2\over m_{K,0}^2}+{4\over9}\log{\Lambda^2\over m_{\eta,0}^2}
  \right)\right]B_0^2m_s^2\;.
%\\ &&
\label{normal}
\eqa
Using the renormalization group equations (\ref{rgeq}) for the
couplings, we find that the effective potential is independent of the
renormalization scale $\Lambda$.
%The quasiparticle masses are given by Eqs.~(\ref{nspec1})--(\ref{nspec2}).
%From these expressions, it is straightforward to see
We note that the renormalized effective potential of Eq.~(\ref{normal}) is independent of the chemical potentials 
$\mu_I$ and $\mu_S$. This independence is a result that we expect will generalize at next-to-next-to-leading order (NNLO) and higher orders. This is due to a general argument, namely the Silver Blaze property, that shows the isospin independence of the eigenvalues of the Dirac operator at finite isospin density (in the normal phase)~\cite{cohen} and consequently the isospin independence of the partition function and resulting thermodynamic quantities. While the original proof in Ref.~\cite{cohen} did not include the strange chemical potential, we expect that it generalizes to systems with both isospin and strange chemical potentials.

\subsection{Pion-condensed phase}
The tree-level contribution to the effective potential is minus
the static Lagrangian given in Eq.~(\ref{statlag1})
\bqa
V_0&=&-f^2B_0(2m\cos\alpha+m_s)-{1\over2}f^2\mu_I^2\sin^2\alpha
%=-f^2\left[m_{\pi,0}^2\left(\cos\alpha-\mbox{$1\over2$}\right)+m_{K,0}^2\right]
%-{1\over2}f^2\mu_I^2\sin^2\alpha
\;.
\label{treepi}
\eqa
The one-loop effective potential is
\bqa
V_{\rm1}&=&{1\over2}\int_p\left[
  E_{\pi^+}+E_{\pi^-}+E_{\pi^0}
+  E_{K^+}+E_{K^-}+E_{K^0}+E_{\bar{K}^0}+E_{\eta^0}
\right]\;.
\eqa
where the energies are given by Eqs.~(\ref{kvasi1})--(\ref{kvasi2}).
%The one-loop contribution to the effective is divergent in the
%ultraviolet and needs renormalization.
The integrals of $E_{\pi^0}$ and $E_{\eta^0}$ can be calculated
analytically in dimensional regularization
using Eq.~(\ref{int1}).
The remaining contributions require a little more work.
Let us consider the contribution from the charged pions.
In order to eliminate the divergences, 
their dispersion relations are expanded in powers of $1/p$ as
\bqa
E_{\pi^+}+E_{\pi^-}=
2p+\frac{2(m_{1}^{2}+m_{2}^{2})+m_{12}^{2}}{4p}
-\frac{8(m_{1}^{4}+m_{2}^{4})+4(m_{1}^{2}+m_{2}^{2})m_{12}^{2}
  +m_{12}^{4}}{64p^{3}}+\dots\;\;\;\;\;\;\;\;
\label{expandp}
\eqa
To this order, the large-$p$ behavior in Eq.~(\ref{expandp}) is the same
as the sum $E_1+E_2$, where 
$E_1=\sqrt{p^2+m_1^2+\mbox{$1\over4$}m_{12}^2}$ and
$E_2=\sqrt{p^2+m_2^2+\mbox{$1\over4$}m_{12}^2}$.
For later convenience we introduce
the masses $\tilde{m}_1^2=m_1^2+{1\over4}m_{12}^2=2B_0m\cos\alpha$,
$\tilde{m}_2^2=m_2^2+{1\over4}m_{12}^2=2B_0m\cos\alpha+\mu_I^2\sin^2\alpha=m_3^2$,
The integral over $E_{\pi^+}+E_{\pi^-}-E_1-E_2$ is convergent
in the ultraviolet
and the subtraction integrals of $E_1$ and $E_2$ can be done analytically in
dimensional regularization. %The result is given in Eq.~(\ref{int1}).
We can then write
\bqa
V_{1,\pi^+}+V_{1,\pi^-}
&=&V_{{\rm 1},\pi^{+}}^{\rm div}+V_{{\rm 1},\pi^{-}}^{\rm div}
+
V_{{\rm 1},\pi^{+}}^{\rm fin}+V_{{\rm 1},\pi^{-}}^{\rm fin}
\eqa
where
\bqa
V_{{\rm 1},\pi^{+}}^{\rm div}+V_{{\rm 1},\pi^{-}}^{\rm div}
&=&{1\over2}\int_p\left[E_1+E_2\right]
\;,\\
V_{{\rm 1},\pi^{+}}^{\rm fin}+V_{{\rm 1},\pi^{-}}^{\rm fin}
&=&\frac{1}{2}\int_{p}\left [E_{\pi^+}+E_{\pi^-}-E_1-E_2\right ]\;.
\eqa
The contributions from the kaons can be calculated analytically as follows.
Consider first the contribution from the charged kaon which is
given by
\bqa
V_{1,K^{+}}+V_{1,K^{-}}
&=&{1\over2}\int_P\log
\left[(P^2+m_4^2)(P^2+m_5^2)+p_0^2m_{45}^2\right]\;,
\eqa
which can be rewritten as
\bqa
V_{1,K^{+}}+V_{1,K^{-}}
&=&{1\over2}\int_P\log\left\{
  \left[P^2+{1\over2}(m_4^2+m_5^2)\right]^2+p_0^2m_{45}^2
  -{1\over4}(m_4^2-m_5^2)^2
\right\}\;.  
\eqa
Since $m_4=m_5$, the last term vanishes and the integrand can be
factorized as
\bqa\nonumber
V_{1,K^{+}}+V_{1,K^{-}}&=&
{1\over2}\int_P\log\left[\left(p_0+{im_{45}\over2}\right)^2+p^2+
  m_4^2+{1\over4}m_{45}^2\right]
\\ &&
\times\left[\left(p_0-{im_{45}\over2}\right)^2+p^2+
m_4^2+{1\over4}m_{45}^2\right]
\;.
\eqa
Shifting integration variables in the two terms,
$p_0\rightarrow p_0\mp{im_{45}\over2}$, the integral
simplifies to
\bqa
V_{1,K^{+}}+V_{1,K^{-}}
&=&\int_P\log\left[P^2+m_4^2+{1\over4}m_{45}^2\right]\;.
\eqa
The contribution from the neutral kaons is obtained simply by replacing
$m_4$ by $m_6$ and $m_{45}$ by $m_{67}$.
Since $\tilde{m}_2^2=m_3^2$ and by defining
$\tilde{m}_4^2=m_4^2+{1\over4}m_{45}^2=m_6^2+{1\over4}m_{67}^2=
B_0(m\cos\alpha+m_s)+{1\over4}\mu_I^2\sin^2\alpha$,
we can write the divergent part of the one-loop contribution as
\bqa\nonumber
V_{\rm 1}^{\rm div}
&=&
-\frac{1}{4(4\pi)^{2}}\left [\frac{1}{\epsilon}+\frac{3}{2} +
  \log\left (\frac{\Lambda^{2}}{{\tilde{m}_1^2}}
  \right ) \right ]\left[2B_0m\cos\alpha\right]^2
%\\ &&\nonumber-\frac{1}{4(4\pi)^{2}}\left [\frac{1}{\epsilon}+\frac{3}{2} +
%  \log\left (\frac{\Lambda^{2}}{{\tilde{m}_2^2}}
%  \right ) \right ]\left[
%  \left(2B_0m\cos\alpha  +
%    \mu_{I}^2\sin^2\alpha\right)^2\right]
\\ &&\nonumber
-\frac{1}{2(4\pi)^{2}}\left [\frac{1}{\epsilon}+\frac{3}{2} +
  \log\left (\frac{\Lambda^{2}}{m_{3}^{2}} \right )\right ]
\left[2B_0m\cos\alpha  +
  \mu_{I}^2\sin^2\alpha\right]^2
\\ \nonumber
&&-\frac{1}{(4\pi)^{2}}\left [\frac{1}{\epsilon}+\frac{3}{2} +
  \log\left (\frac{\Lambda^{2}}{\tilde{m}_4^{2}} \right )\right ]
\left[B_0(m\cos\alpha+m_s)+\mbox{$1\over4$}\mu_{I}^{2}\sin^{2}\alpha\right]^2
\\ 
&& %\nonumber
-\frac{1}{4(4\pi)^{2}}\left [\frac{1}{\epsilon}+\frac{3}{2} +
  \log\left (\frac{\Lambda^{2}}{m_{8}^{2}} \right )\right ]
\left[{2B_0(m\cos\alpha+2m_s)\over3}\right]^2
%\\ &&
%+V_{\rm eff,{\pi^{\pm}}}^{\rm fin}
%+V_{\rm eff,{K^{\pm}}}^{\rm fin}
%+V_{\rm eff,{K^{0}}}^{\rm fin}
\label{divpi}
\;,
\eqa
The static part of the Lagrangian ${\cal L}_4$ as a function of $\alpha$ is
%\begin{widetext}
  \bqa\nonumber
  V_{1}^{\rm static}
  &=& -(4L_1+4L_2+2L_3)\mu_I^4\sin^4\alpha
  -8L_4B_0(2m\cos\alpha+m_s)\mu_I^2\sin^2\alpha
  \\ &&\nonumber
  -8L_5B_0m\mu_I^2
  \cos\alpha\sin^2\alpha
  -16L_6B_0^2(2m\cos\alpha+m_s)^2  
\\ &&
  -  8L_8B_0^2(2m^2\cos2\alpha+m_s^2)
  -4H_2B_0^2(2m^2+m_s^2)\;.
\label{statpi}
  \eqa
  The renormalized one-loop effective potential
  $V_{\rm eff}=V_0+V_1+V_1^{\rm static}$ is given by the sum of
  Eqs.~(\ref{treepi}), ~(\ref{divpi}), and ~(\ref{statpi})
  then reads
\bqa\nonumber
V_{\rm eff}&=&-f^2B_0(2m\cos\alpha+m_s)
-{1\over2}f^2\mu_I^2\sin^2\alpha
-  (4{L}_1^r+4{L}_2^r+2{L}_3^r)\mu_I^4\sin^4\alpha
\\ && \nonumber
-8{L}_4^rB_0(2m\cos\alpha+m_s)\mu_I^2\sin^2\alpha
-  8{L}_5^rB_0m\mu_I^2\cos\alpha\sin^2\alpha
\\ && \nonumber
-  16{L}_6^rB_0^2(2m\cos\alpha+m_s)^2
    -8{L}_8^rB_0^2(2m^2\cos2\alpha+m_s^2)-4{H}_2^rB_0^2(2m^2+m_s^2)
    \\ \nonumber
&&-\frac{1}{4(4\pi)^{2}}\left [\frac{1}{2} +
  \log\left (\frac{\Lambda^{2}}{{\tilde{m}_1^2}}
  \right ) \right ]\left[2B_0m\cos\alpha\right]^2    \\ \nonumber
&&-\frac{1}{2(4\pi)^{2}}\left [
  \frac{1}{2} +
  \log\left (\frac{\Lambda^{2}}{m_{3}^{2}} \right )\right ]
\left[2B_0m\cos\alpha  +%\mbox{$1\over2$}
  \mu_{I}^2\sin^2\alpha\right]^2
%\\ \nonumber
%&&-\frac{1}{4(4\pi)^{2}}\left [\frac{1}{2} +
%  \log\left (\frac{\Lambda^{2}}{{m_3^2}}
%  \right ) \right ]\left[
%  2B_0m\cos\alpha  +%\mbox{$1\over2$}
%    \mu_{I}^2\sin^2\alpha
%  +4G^2m^2\cos^2\alpha
%\right]^2
\\ \nonumber
&&-\frac{1}{(4\pi)^{2}}\left [\frac{1}{2} +
  \log\left (\frac{\Lambda^{2}}{\tilde{m}_4^{2}} \right )\right ]
\left[B_0(m\cos\alpha+m_s)+\mbox{$1\over4$}\mu_{I}^{2}\sin^{2}\alpha\right]^2
\\ 
&&-\frac{1}{4(4\pi)^{2}}\left [\frac{1}{2} +
  \log\left (\frac{\Lambda^{2}}{m_{8}^{2}} \right )\right ]
\left[{2B_0(m\cos\alpha+2m_s)\over3}\right]^2
%\\ &&
+V_{\rm 1,{\pi^{+}}}^{\rm fin}
+V_{\rm 1,{\pi^{-}}}^{\rm fin}
%+V_{\rm eff,{K^{\pm}}}^{\rm fin}+V_{\rm eff,{K^{0}}}^{\rm fin}
\;.
\label{renpi}
\eqa
Again, it can be verified that the NLO effective potential is independent
of the scale $\Lambda$. It is also explicitly independent of the
strangeness chemical potential $\mu_S$.

\subsection{Charged kaon-condensed phase}
The tree-level contribution to the effective potential is
\bqa
V_0&=&-f^2B_0\left[m+(m+m_s)\cos\alpha)\right]
-{1\over2}f^2\left(\mbox{$1\over2$}\mu_I+\mu_S\right)^2\sin^2\alpha\;.
\eqa
The one-loop effective potential is
\bqa
V_{\rm1}&=&{1\over2}\int_p\left[
  E_{\pi^+}+E_{\pi^-}+E_{\pi^0}
+  E_{K^+}+E_{K^-}+E_{K^0}+E_{\bar{K}^0}+E_{\eta^0}
\right]\;.
\eqa
The contributions from $\pi^{\pm}$, $K^{\pm}$, $K^{0}$ and $\bar{K}^{0}$
can be treated as in the previous section and it is only
the terms $V_{1,K^{\pm}}$ that require a subtraction term.
The relevant masses are defined as
\bqa
\tilde{m}_1^2&=&m_1^2+{1\over4}m_{12}^2=
{1\over2}B_0[3m-m_s+(m+m_s)\cos\alpha]
+{1\over4}
\left(\mbox{$1\over2$}\mu_I+\mu_S\right)^2\sin^2\alpha
\;,\\
\tilde{m}_4^2&=&
m_4^2+{1\over4}m_{45}^2=
B_0(m+m_s)\cos\alpha
\;,\\
\label{m4tilde}
\tilde{m}_5^2&=&
m_5^2+{1\over4}m_{45}^2=
B_0(m+m_s)\cos\alpha
+\left(\mbox{$1\over2$}\mu_I+\mu_S\right)^2\sin^2\alpha
\label{m5tilde}\;,\\
\tilde{m}_6^2&=&
m_6^2+{1\over4}m_{67}^2=
\mbox{$1\over2$}B_0(m+m_s)(1+\cos\alpha)
+{1\over4}\left(\mbox{$1\over2$}\mu_I+\mu_S\right)^2\sin^2\alpha
\;.
\eqa
The contribution from the
mixed $\pi^0$ and $\eta^0$ is given by 
\bqa\nonumber
V_{1,\pi^0}+V_{1,\eta^0}
&=&%{1\over2}\log D_{38}^{-1}
{1\over2}\int_P\log\left[(P^2+m_3^2)(P^2+m_8^2)-m_{38}^4\right]
\\ &=&
{1\over2}\int_P\log\left[P^2+\tilde{m}_3^2]+\log[P^2+\tilde{m}_8^2\right]
\;,
\eqa
where the new masses are defined as
\bqa
\tilde{m}_{3,8}^2&=&
{1\over2}\left[m_3^2+m_8^2\pm{\sqrt{(m_3^2-m_8^2)^2+4m_{38}^4}}\right]\;.
\eqa
This yields
\bqa\nonumber
V_{\rm 1}
&=&
-\frac{1}{2(4\pi)^{2}}\left[\frac{1}{\epsilon}+{3\over2}
  \log\left (\frac{\Lambda^{2}}{\tilde{m}_{1}^{2}} \right )\right ]
\left\{\mbox{$1\over2$}B_0[(3m-m_s+(m+m_s)\cos\alpha]
\right.\\ &&\nonumber\left.
  +\mbox{$1\over4$}
  \left(\mbox{$1\over2$}\mu_I+\mu_S\right)^2\sin^2\alpha\right\}^2
\\ \nonumber
&&-\frac{1}{4(4\pi)^{2}}\left [\frac{1}{\epsilon}+{3\over2}+
  \log\left (\frac{\Lambda^{2}}{{\tilde{m}_4^2}}
  \right ) \right ]\left[B_0^2(m+m_s)^2\cos^2\alpha
\right]
\\ \nonumber
&&-\frac{1}{4(4\pi)^{2}}\left [\frac{1}{\epsilon}+{3\over2}+
  \log\left (\frac{\Lambda^{2}}{{\tilde{m}_5^2}}
  \right ) \right ]\left[B_0(m+m_s)\cos\alpha  +  
    \left(\mbox{$1\over2$}
    \mu_I+\mu_S\right)^2\sin^2\alpha\right]^2
\\ \nonumber
&&-\frac{1}{2(4\pi)^{2}}\left [\frac{1}{\epsilon}+{3\over2}+
  \log\left (\frac{\Lambda^{2}}{\tilde{m}_6^{2}} \right )\right ]
\left[\mbox{$1\over2$}B_0(m+m_s)(1+\cos\alpha)  +
  \mbox{$1\over4$}
  \left(\mbox{$1\over2$}\mu_I+\mu_S\right)^2\sin^2\alpha
\right]^2
\\ 
&&-\frac{1}{4(4\pi)^{2}}\left [\frac{1}{\epsilon}+\frac{3}{2} +
  \log\left (\frac{\Lambda^{2}}{\tilde{m}_{3}^{2}} \right )\right]\tilde{m}_3^4
-\frac{1}{4(4\pi)^{2}}\left [\frac{1}{\epsilon}+\frac{3}{2} +
  \log\left (\frac{\Lambda^{2}}{\tilde{m}_{8}^{2}} \right )\right]\tilde{m}_8^4
\eqa
The static part of the Lagrangian ${\cal L}_4$ as a function of $\alpha$
is
\bqa\nonumber
V_1^{\rm static}
&=&
-  (4L_1+4L_2+2L_3)\left(\mbox{$1\over2$}\mu_I+\mu_S\right)^4\sin^4\alpha
\\ && \nonumber
-8L_4B_0[m+(m+m_s)\cos\alpha]
  \left(\mbox{$1\over2$}
    \mu_I+\mu_S\right)^2\sin^2\alpha
\\ \nonumber&&
-4L_5B_0(m+m_s)\left(\mbox{$1\over2$}\mu_I+\mu_S\right)^{{2}}\cos\alpha
\sin^2\alpha
-  16L_6B_0^2[m+(m+m_s)\cos\alpha]^2
  \label{lagstat}
\\ &&
-4L_8B_0^2(3m^2 -2mm_s+m_s^2+(m+m_s)^2\cos2\alpha)-4H_2B_0^2(2m^2+m_s^2)\;.
\eqa
After renormalization, the effective potential is
\bqa\nonumber
V_{\rm eff}
&=&
-f^2B_0\left[m+(m+m_s)\cos\alpha)\right]
-\mbox{$1\over2$}f^2\left(\mbox{$1\over2$}\mu_I+\mu_S\right)^2\sin^2\alpha
\\ && \nonumber
-(4{L}_1^r+4{L}_2^r+2{L}_3^r)
\left(\mbox{$1\over2$}\mu_I+\mu_S\right)^4\sin^4\alpha
\\ &&\nonumber
-8{L}_4^rB_0[m+(m+m_s)\cos\alpha]
  \left(\mbox{$1\over2$}
    \mu_I+\mu_S\right)^2\sin^2\alpha
\\ \nonumber&&
-4{L}_5^rB_0(m+m_s)\left(\mbox{$1\over2$}\mu_I+\mu_S\right)^2\cos\alpha
\sin^2\alpha
-  16{L}_6^rB_0[m+(m+m_s)\cos\alpha]^2
\\ && \nonumber
-4{L}_8^rB_0^2(3m^2 -2mm_s+m_s^2+(m+m_s)^2\cos2\alpha)-4{H}_2^rB_0^2(2m^2+m_s^2)
\\ && \nonumber
-\frac{1}{2(4\pi)^{2}}\left[{1\over2}+
  \log\left (\frac{\Lambda^{2}}{\tilde{m}_{1}^{2}} \right )\right ]
\Big\{\mbox{$1\over2$}B_0[(3m-m_s+(m+m_s)\cos\alpha]
%\right.\\ &&\left.\nonumber
%\\ && \nonumber
+\mbox{$1\over4$}
  \left(\mbox{$1\over2$}\mu_I+\mu_S\right)^2\sin^2\alpha\Big\}^2
\\ \nonumber
&&-\frac{1}{4(4\pi)^{2}}\left [{1\over2}+
  \log\left (\frac{\Lambda^{2}}{{\tilde{m}_4^2}}
  \right ) \right ]\left[
B_0^2(m+m_s)^2\cos^2\alpha\right]
\\ \nonumber
&&-\frac{1}{4(4\pi)^{2}}\left [{1\over2}+
  \log\left (\frac{\Lambda^{2}}{{\tilde{m}_5^2}}
  \right ) \right ]\left[B_0(m+m_s)\cos\alpha  +  
%    \mbox{$1\over2$}
    \left(\mbox{$1\over2$}\mu_I+\mu_S\right)^2
    \sin^2\alpha\right]^2
\\ \nonumber
&&-\frac{1}{2(4\pi)^{2}}\left [
  {1\over2}+
  \log\left (\frac{\Lambda^{2}}{\tilde{m}_6^{2}} \right )\right ]
\left[\mbox{$1\over2$}B_0(m+m_s)(1+\cos\alpha)  +
  \mbox{$1\over4$}
  \left(\mbox{$1\over2$}\mu_I+\mu_S\right)^2\sin^2\alpha
\right]^2
\\ \nonumber
&&-\frac{1}{4(4\pi)^{2}}\left [  \frac{1}{2} +
  \log\left (\frac{\Lambda^{2}}{\tilde{m}_{3}^{2}} \right )\right]\tilde{m}_3^4
-\frac{1}{4(4\pi)^{2}}\left [
  \frac{1}{2} +
  \log\left (\frac{\Lambda^{2}}{\tilde{m}_{8}^{2}} \right )\right]\tilde{m}_8^4
%\\ && 
+V_{1,{K^{+}}}^{\rm fin}
+V_{1,{K^{-}}}^{\rm fin}\;,
\\ &&
\label{nlok}
\eqa
where the subtraction terms and energies are defined by
\bqa
V_{1,{K^{+}}}^{\rm fin}
+V_{1,{K^{-}}}^{\rm fin}&=&
{1\over2}\int_p\left[E_{K^{+}}+E_{K^-}-E_4-E_5
\right]\;,
\\ 
E_{4,5}&=&\sqrt{p^2+\tilde{m}_{4,5}^2}\;,
\eqa
with $\tilde{m}_{4,5}$ given by Eqs.~(\ref{m4tilde})--(\ref{m4tilde}).

The effective potential depends only on the combination
$|\mbox{$1\over2$}\mu_I+\mu_S|$ as is evident by inspection.
Using the expressions for the running couplings, Eq.~(\ref{rgeq}),
the scale dependence in the final results for the effective potential,
Eqs.~(\ref{normal}), ~(\ref{renpi}), and ~(\ref{nlok}) cancels.

%The effective potential depends only on the combination
%$\mbox{$1\over2$}\mu_I+\mu_S$.

\section{Thermodynamic functions}
In this section, we derive various thermodynamic functions from the
effective potential. We will focus on the pion-condensed phase
since we are interested in comparing our results with lattice simulations.

\subsection{Pion-condensed phase}
The pressure $P$ is given by $-V_{\rm eff}$.
In the pion-condensed phase, we get from Eq.~(\ref{renpi})
\bqa
    \nonumber
P&=&f^2B_0(2m\cos\alpha+m_s)+{1\over2}f^2\mu_I^2\sin^2\alpha
\\ && \nonumber
+\left[4{L}_1^r+4{L}_2^r+2{L}_3^r
%\right.\\ &&\left.\nonumber
  +{1\over16(4\pi)^2}\left({9\over2}+8\log{\Lambda^2\over m_3^2}
%\right.\right.\\ && \nonumber\left.\left.
  +\log{\Lambda^2\over\tilde{m}_4^2}\right)
            \right]\mu_I^4\sin^4\alpha
\\ && \nonumber
+\left[8{L}_4^{r}
  +{1\over2(4\pi)^2}\left({1\over2}+\log{\Lambda^2\over\tilde{m}_4^2}\right)
\right]
%\\&&\times \nonumber
                  B_0(2m\cos\alpha+m_s)\mu_I^2\sin^2\alpha
\\ && \nonumber
+\left[8{L}_5^{r}
  +{1\over2(4\pi)^2}\left({3\over2}+4\log{\Lambda^2\over m_3^2}
 -\log{\Lambda^2\over\tilde{m}_4^2}\right)     \right]
%  \\&&\times \nonumber
  B_0m\mu_I^2\cos\alpha\sin^2\alpha
\\ && \nonumber
+\left[
  16L_6^r+8L_8^r+4H_2^r
  +{1\over(4\pi)^2}\left({13\over18}  +\log{\Lambda^2\over\tilde{m}_4^2}
%\right.\right.\\ && \nonumber\left.\left.
+{4\over9}\log{\Lambda^2\over m_8^2}\right)\right]B_0^2m_s^2
\\ && \nonumber
+\left[  64L_6^r  +{1\over(4\pi)^2}\left({11\over9}
    +2\log{\Lambda^2\over\tilde{m}_4^2}
+{4\over9}\log{\Lambda^2\over m_8^2}  \right)\right]
%\\&& \times \nonumber
     B_0^2mm_s\cos\alpha
\\ && \nonumber
+\left[
  64L_6^r+16L_8^r+8H_2^r    +{1\over(4\pi)^2}\left({37\over18}
+\log{\Lambda^2\over\tilde{m}_1^2}+
\right.\right. \\ &&\left.\left. \nonumber
     +2\log{\Lambda^2\over m_3^2}+    \log{\Lambda^2\over\tilde{m}_4^2}
+{1\over9}\log{\Lambda^2\over m_8^2}  \right)\right]B_0^2m^2\cos^2\alpha
-\left[16L_8^r-8H_2^r\right]B_0^2m^2\sin^2\alpha
\\ &&
-V_{\rm 1,{\pi^{+}}}^{\rm fin}
-V_{\rm 1,{\pi^{-}}}^{\rm fin}\;,
\label{renpi2}
\eqa
The isospin density is given by
\bqa\nonumber
n_I&=&-{\partial V_{\rm eff}\over\partial\mu_I}\\ \nonumber
&=& \nonumber
f^2\mu_I\sin^2\alpha
+\left[16L_1^r+16L_2^r+8L_3^r
  +{1\over{4(4\pi)^2}}\left({8}\log{\Lambda^2\over m_3^2}
+\log{\Lambda^2\over\tilde{m}_4^2}\right)\right]\mu_I^3\sin^4\alpha
\\ && \nonumber
+\left[{16}L_4^r%+{16}L_5^r
  +{{1}\over{(4\pi)^2}}
\log{\Lambda^2\over\tilde{m}_4^2}\right]
B_0(2m\cos\alpha+m_s)\mu_I\sin^2\alpha
\\ &&
+\left[{16}L_5^r+{1\over{(4\pi)^2}}\left(
    4\log{\Lambda^2\over{m}_3^2}
      -\log{\Lambda^2\over\tilde{m}_4^2}\right)
\right]
B_0m\mu_I\cos\alpha\sin^2\alpha
-{\partial V_{\rm 1,{\pi^{+}}}^{\rm fin}\over\partial\mu_I}
-{\partial V_{\rm 1,{\pi^{-}}}^{\rm fin}\over\partial\mu_I}
\;.
\label{isos}
\eqa
The energy density is given by
\bqa
\epsilon&=&-P+\mu_in_i\;.
\eqa
where $n_i=-{\partial V_{\rm eff}\over\partial\mu_i}$ is the
charge density associated with the chemical potential $\mu_i$. In the 
pion-condensed phase it takes the following form 
\bqa
\epsilon&=&-P+\mu_In_I\;.
\eqa
since the effective potential is independent of $\mu_S=0$ in this phase.
Using Eqs.~(\ref{renpi}) and~(\ref{isos}),
we find the following energy density
\bqa\nonumber
\epsilon&=&
-f^2B_0(2m\cos\alpha+m_s)
+{1\over2}f^2\mu_I\sin^2\alpha
\\ &&\nonumber
+\left[12L_1^r+12L_2^r+6L_3^r+{1\over{(4\pi)^2}}
  \left(-{9\over32}+{{\frac{3}{2}}}\log{\Lambda^2\over m_3^2}
+{3\over16}\log{\Lambda^2\over\tilde{m}_4^2}\right)\right]\mu_I^4\sin^4\alpha
\\ && \nonumber
+\left[{8}L_4^r+{1\over2{(4\pi)^2}}
  \left({-\frac{1}{2}}
    +\log{\Lambda^2\over\tilde{m}_4^2}\right)\right]
B_0(2m\cos\alpha+m_s)\mu_I^2\sin^2\alpha
\\ && \nonumber
+\left[{8}L_5^r+{1\over{2(4\pi)^2}}\left({-{3\over 2}}+
    4\log{\Lambda^2\over{m}_3^2}-
  \log{\Lambda^2\over\tilde{m}_4^2}\right)
\right]
B_0m\mu_I^2\cos\alpha\sin^2\alpha
\\ \nonumber&&
-
%\right.\\ && \nonumber\left.
\left[
  16L_6^r+8L_8^r+4H_2^r
  +{1\over(4\pi)^2}\left({13\over18}
  +\log{\Lambda^2\over\tilde{m}_4^2}
    +{4\over9}\log{\Lambda^2\over m_8^2}
    \right)\right]B_0^2m_s^2
\\ && \nonumber
-\left[
  64L_6^r
  +{1\over(4\pi)^2}\left({11\over9}
    +2\log{\Lambda^2\over\tilde{m}_4^2}
    +{4\over9}\log{\Lambda^2\over m_8^2}
  \right)\right]
%\\&& \times \nonumber
     B_0^2mm_s\cos\alpha
\\ && \nonumber
-\left[  64L_6^r+16L_8^r+8H_2^r
     +{1\over(4\pi)^2}\left({37\over18}
+\log{\Lambda^2\over\tilde{m}_1^2}
+2\log{\Lambda^2\over m_3^2}
+    \log{\Lambda^2\over\tilde{m}_4^2}
%\right.\right. \\ &&\left.\left. 
+{1\over9}\log{\Lambda^2\over m_8^2}\right)\right]
\\ && \times B_0^2m^2\cos^2\alpha
+V^{\rm fin}_{\rm 1,{\pi^{+}}}+V^{\rm fin}_{\rm 1,{\pi^{-}}}
-\mu_I{\partial V_{\rm 1,{\pi^{+}}}^{\rm fin}\over\partial\mu_I}
-\mu_I{\partial V_{\rm 1,{\pi^{-}}}^{\rm fin}\over\partial\mu_I}
\label{eos0}
\;,
\eqa
\subsection{Large-$m_s$ limit}
\label{mapping}
We are interested in the large-$m_s$ limit of our three-flavor results for
thermodynamic quantities. In this limit, general effective field theory
arguments tell us that the mesonic degrees of freedom containing the $s$-quark
decouple. Thus one should recover the two-flavor results of
Ref.~\cite{us} with modified couplings.
The modified couplings then contain the loop effects from
integrating out kaons and the eta.

The one-loop expressions for the pion-decay constant
and the light-quark condensate in the vacuum 
are given by~\cite{gasser2}
\bqa
\label{fpi0}
f_{\pi}^2
&=&f^2\left[1
  +\left(8{L}_4^r
+8L_5^r+{2\over(4\pi)^2}\log{\Lambda^2\over m_{\pi,0}^2}
\right){m_{\pi,0}^2\over f^2}
+\left(16L_4^r+{1\over(4\pi)^2}
  \log{\Lambda^2\over m_{K,0}^2}
  \right){m_{K,0}^2\over f^2}
\right]
\\  \nonumber
\langle\bar{\psi}\psi\rangle&=&-f^2B_0\left[
  1+
  \left(16L_6^r+4L_8^r+4H_2^r
+  {3\over2(4\pi)^2}\log{\Lambda^2\over m_{\pi,0}^2}
\right){m_{\pi,0}^2\over f^2}
\right.\\ &&\left.
  +\left(
    32L_6^r+{1\over(4\pi)^2}
    \log{\Lambda^2\over m_{K,0}^2}
  \right){m_{K,0}^2\over f^2}
  +{m_{\eta,0}^2\over6(4\pi)^2f^2}\log{\Lambda^2\over m_{\eta,0}^2}
\right]\;.
\eqa
The loop corrections involve pions, kaons, and etas. Integrating out
the $s$-quark corresponds to setting $m=0$ or ignoring the pionic loop
corrections. This yields
\bqa
\tilde{f}^2&=&f^2\left[1+
  \left(16L_4^r
%\right.\right.\\ &&\left.\left.    
    +{1\over(4\pi)^2}\log{\Lambda^2\over \tilde{m}_{K,0\textrm{}}^2}\right)
{\tilde{m}_{K,0}^2\over f^2}
\right]\;,
\label{rel5}
\\
\langle\bar{\psi}\psi\rangle&=&-f^2B_0\left[1
  +\left(
    32L_6^r+{1\over(4\pi)^2}
    \log{\Lambda^2\over \tilde{m}_{K,0}^2}
  \right){\tilde{m}_{K,0}^2\over f^2}
  +{\tilde{m}_{\eta,0}^2\over6(4\pi)^2f^2}\log{\Lambda^2\over \tilde{m}_{\eta,0}^2}
\right]\;,
\eqa
where the masses are 
$\tilde{m}_{K,0}^2=B_0m_s$ and $\tilde{m}_{\eta,0}^2={4B_0m_s\over3}$.
Defining $\tilde{B}_0$ via
$\langle\bar{\psi}\psi\rangle=-\tilde{f}^2\tilde{B}_0$
yields
\bqa
\tilde{B}_0&=&B_0\left[
  1-\left(16L_4^r-32L_6^r\right)
    {\tilde{m}_{K,0}^2\over f^2}
    +{\tilde{m}_{\eta,0}^2\over6(4\pi)^2f^2}\log{\Lambda^2\over \tilde{m}_{\eta,0}^2}
\right]\;.
\label{rel6}
\eqa
Using the renormalization group equations for $L_4^r$ and $L_6^r$,
one verifies that Eqs.~(\ref{rel5}) and~(\ref{rel6})
are independent of the scale $\Lambda$.
Moreover, in Ref.~\cite{gasser2}, the authors derived the
relations among the renormalized couplings in two - and three-flavor
$\chi$PT. The relevant relations are
\bqa%\nonumber
l_1^r&=&4L_1^r+2L_3^r
+{1\over48(4\pi)^2}
\left[\log{\Lambda^2\over\tilde{m}_{K,0\textrm{}}^2}-1\right]\;,
\label{rel0} \\
l_2^r&=&4L_2^r
+{1\over24}{1\over(4\pi)^2}
\left[\log{\Lambda^2\over\tilde{m}_{K,0\textrm{}}^2}-1\right]\;,
\label{rel1}
\\ %\nonumber
l_3^r&=&-8L_4^r-4L_5^r+16L_6^r+8L_8^r
+{1\over36(4\pi)^2}
\left[\log{\Lambda^2\over\tilde{m}_{\eta,0\textrm{}}^2}-1\right]\;,
\\
l_4^r&=&8L_4^r+4L_5^r+{1\over4(4\pi)^2}\left[
  \log{\Lambda^2\over\tilde{m}_{K,0\textrm{}}^2}-1\right]\;,
\\
h_1^r&=&8L_4^r+4L_5^r-4L_8^r+2H_2^r
+{1\over4(4\pi)^2}\left[
  \log{\Lambda^2\over\tilde{m}_{K,0\textrm{}}^2}-1\right]\;.
\label{rel4}
\eqa
The relations between the renormalized couplings $l_i^r,h_i^r$ and the
low-energy constants $\bar{l}_i, \bar{h}_i$ in two-flavor $\chi$PT are
\bqa
l_i^r(\Lambda)&=&{\gamma_i\over2(4\pi)^2}
\left[\bar{l}_i+\log{2B_0m\over\Lambda^2}\right]\;,
\hspace{1cm}
h_i^r(\Lambda)={\delta_i\over2(4\pi)^2}
\left[\bar{h}_i+\log{2B_0m\over\Lambda^2}\right]\;,
\label{lirlbare}
\eqa
where $\gamma_1={1\over3}$, $\gamma_2={2\over3}$, $\gamma_3=-{1\over2}$, 
$\gamma_4=2$, and $\delta_1=2$~\cite{gasser1}.
Using the renormalization group equations
for renormalized couplings, one finds that the $\Lambda$-dependence
are the same on the left - and right-hand
sides of Eqs.~(\ref{rel0})--(\ref{rel4}).

The large-$m_{s}$ limit of Eqs.~(\ref{renpi2}),~(\ref{isos}), and~(\ref{eos0})
are then obtained as follows. We expand them in powers of
$1/m_s$, express the result using Eqs.~(\ref{rel5}), and
(\ref{rel6})--(\ref{lirlbare}). The pressure is
\bqa\nonumber
P&=&2\tilde{f}^{2}\tilde{B}_0m\cos\alpha+\frac{1}{2}\tilde{f}^{2}\mu_{I}^{2}
\sin^{2}\alpha
+{4\over(4\pi)^2}\left[-\bar{h}_1+\bar{l}_1\right]B_0^2m^2
\\ && \nonumber
+\frac{1}{(4\pi)^{2}}\left [\frac{3}{2}-\bar{l}_{3}
  +{4}\bar{l}_{4}
  +\log\left({2B_0m\over \tilde{m}_1^2}\right)
  +  {2}\log{2B_0m\over m_3^2}\right ]B_0^2m^{2}\cos^{2}\alpha
\\&& \nonumber
+\frac{1}{(4\pi)^{2}}
\left [{1\over2}+\bar{l}_{4}
  +  \log{2B_0m\over m_3^2}\right ]
m^{2}\mu_{I}^{2}\cos\alpha\sin^{2}\alpha
\\ &&
+\frac{1}{2(4\pi)^{2}}\left [{1\over2}
  +\frac{1}{3}\bar{l}_{1}+\frac{2}{3}\bar{l}_{2}
  +  \log{2B_0m\over m_3^2}\right ]\mu_{I}^{4}\sin^{4}\alpha
-V_{{\rm 1},\pi^+}^{\rm fin}-V_{{\rm 1},\pi^-}^{\rm fin}
\;,
\label{effpotnlo}
\eqa
the isospin density is
\bqa\nonumber
n_{I}&=& \nonumber
\tilde{f}^2\mu_I\sin^2\alpha
+{2\over(4\pi)^2}\left[%{-2m_1^2+m_3^2+2m_2^2\over2m_3^2}+
  \bar{l}_4+\log{2B_0m\over m_3^2}\right]
%\\&& \nonumber\times
2B_0m\mu_I\cos\alpha\sin^2\alpha
\\ 
&&+{2\over(4\pi)^2}\left[%{-m_1^2+m_2^2+m_3^2\over2m_3^2}+
  {1\over3}\bar{l}_1+{2\over3}\bar{l}_2+
  \log{2B_0m\over m_3^2}\right]\mu_I^3\sin^4\alpha
%\\&& %\nonumber
-\mu_I{\partial V_{{\rm 1},\pi^+}^{\rm fin}
  \over\partial\mu_I}
-\mu_I{\partial V_{{\rm 1},\pi^-}^{\rm fin}
  \over\partial\mu_I}\;,
\eqa
and the energy density is
\bqa\nonumber
\epsilon&=& \nonumber
-2\tilde{f}^2\tilde{B}_0m\cos\alpha
+{1\over2}\tilde{f}^2\mu_I^2\sin^2\alpha
\\ &&\nonumber
-{4\over(4\pi)^2}\left[-\bar{h}_1+\bar{l}_1\right]B_0^2m^2
-\frac{1}{(4\pi)^{2}}\left [\frac{3}{2}-\bar{l}_{3}
  +{4}\bar{l}_{4}
  +\log\left({2B_0m\over\tilde{m}_1^2}\right)
%\right.\\ && \left.\nonumber
+  {2}\log\left({2B_0m\over m_3^2}\right)
\right ]B_0^2m^{2}\cos^{2}\alpha
\\ && \nonumber
-{1\over(4\pi)^2}\left[{1\over2}-\bar{l}_4
  -\log{2B_0m\over m_3^2}\right]
2B_0m\mu_I^2\cos\alpha\sin^2\alpha
\\ \nonumber
&&-{1\over2(4\pi)^2}\left[{1\over2}
  -\bar{l}_1-2\bar{l}_2-
  3\log{2B_0m\over m_3^2}\right]
\mu_I^4\sin^4\alpha
%\\&&
+V_{{\rm 1},\pi^+}^{\rm fin}+V_{{\rm 1},\pi^-}^{\rm fin}
-\mu_I{\partial V_{{\rm 1},\pi^+}^{\rm fin}
  \over\partial\mu_I}
-\mu_I{\partial V_{{\rm 1},\pi^-}^{\rm fin}
  \over\partial\mu_I}\;.
\\ &&
\eqa
Up to different notation ($2B_0m\rightarrow m^2$),
the results for $P$, $n_I$, and $\epsilon$ are of the same form as
the two-flavor results derived in~\cite{us} with renormalized parameters
$\tilde{B}_0$ and $\tilde{f}$. (In two-flavor $\chi$PT $m$ is the tree level pion mass.~\cite{us})

\section{Results and discussion}
In this section, we calculate and study the (tree-level) quasiparticle masses,
isospin density, pressure and the equation of state.
In order to evaluate these quantities, we need the numerical values
of the low-energy constants ($L_i$) as well as
the meson masses and decay constants. The low-energy constants
have been determined
experimentally, with the following values and uncertainties
at the scale
$\mu=m_{\rho}$, where $\Lambda^2=4\pi e^{-\gamma_E}\mu^2$~\cite{bijnensreview}, where $m_{\rho}$ is the mass of the $\rho$ meson,
\begin{align}
{L}_{1}^r&=(1.0\pm 0.1)\times10^{-3}\;,&
{L}_{2}^r&=(1.6\pm 0.2)\times10^{-3}\;,\\
{L}_{3}^r&=(-3.8\pm 0.3)\times10^{-3}\;,
&{L}_{4}^r&=(0.0 \pm 0.3)\times10^{-3}\;,
\\
{L}_{5}^r&=(1.2 \pm 0.1)\times10^{-3}\;
&{L}_{6}^r&=(0.0 \pm 0.4)\times10^{-3}\;,\\
{L}_{7}^r&=(-0.4 \pm 0.2)\times10^{-3}\;
&{L}_{8}^r&=(0.5 \pm 0.2)\times10^{-3}\;.
\label{LECs}
\end{align}
Since we are mainly interested in comparing our results to the
predictions of the lattice simulations in Refs.~\cite{gergy1},
we will use their values for the pion and kaon masses as well as the pion and kaon decay constants. With uncertainties, they are given
by~\cite{private}
\begin{align}
m_{\pi}&=131\pm3\text{MeV}\;,&
m_{K}=481\pm10\text{MeV}\;,
\\ 
  f_{\pi}&={128\pm3\over\sqrt{2}}           \text{MeV}\;,
&  f_{K}={150\pm3\over\sqrt{2}}           \text{MeV}\;.
\label{latticeval}
\end{align}
These uncertainties (in the masses and decay constants) arise due to lattice discretization errors and consequently differ slightly from their experimental values.
Since we have three parameters in the Lagrangian, $B_0m$, $B_0m_s$, and
$f$, we need to pick three observables from the set above, and
we choose $m_{\pi}$, $m_K$, and $f_{\pi}$.

The relevant meson masses and 
the pion decay constants at one-loop
are given by Eqs.~(\ref{mpi}), ~(\ref{mk}), and~(\ref{fpi})
in terms of the parameters $B_0m$, $B_0m_s$ and $f$ at next-to-leading
order. Using the lattice values given above, we
can solve for $B_0m$, $B_0m_s$, and $f$. This yields
\begin{align}
\label{treevals}
  f^{\rm cen}&=75.16\;{\rm MeV}\;,
  f^{\rm low}=79.88\;{\rm MeV}\;,
  f^{\rm high}=70.44\;{\rm MeV}\;,\\
m_{\pi,\rm tree}^{\rm low}&=148.45\;{\rm MeV}\;,
m_{\pi,\rm tree}^{\rm cen}=131.28\;{\rm MeV}\;,
m_{\pi,\rm tree}^{\rm high}=115.93\;{\rm MeV}\;,\\
m_{K,\rm tree}^{\rm cen}&=520.65\;{\rm MeV}\;,
m_{K,\rm tree}^{\rm low}=617.35\;{\rm MeV}\;,
m_{K,\rm tree}^{\rm high}=437.84\;{\rm MeV}\;,
\end{align}
where the subscripts indicate that the values correspond to the
central, minimum, and maximum values of the low-energy constants.
Using the one-loop $\chi$PT expression for the
$f_K$, Eq.~(\ref{fk}),
we find $f_K=113.9$ MeV for the central values, 
which is off by approximately 7\% compared to the lattice value
of $f_K={150\over\sqrt{2}}=106.1$ MeV. The uncertainties in the LECs, $L_{i}^{r}$, the pion mass, $m_{\pi}$, the pion decay constant, $f_{\pi}$, and the kaon mass, $m_{K}$, lead to uncertainties in $B_{0}m$, $B_{0}m_{s}$ and $f$. These uncertainties are dominated by the uncertainties in the LECs with the uncertainty in the lattice parameters contributing the least. Additionally, it turns out that the lowest values of LECs calculated after including the LEC uncertainty leads to unphysical values of the $\eta$ mass. As such we were forced to choose the lowest values of the LECs using 0.46 times the uncertainties leading to the results in Eq.~(\ref{treevals}). 

\begin{figure}[htb]
\centering  \includegraphics[width=0.45\textwidth]{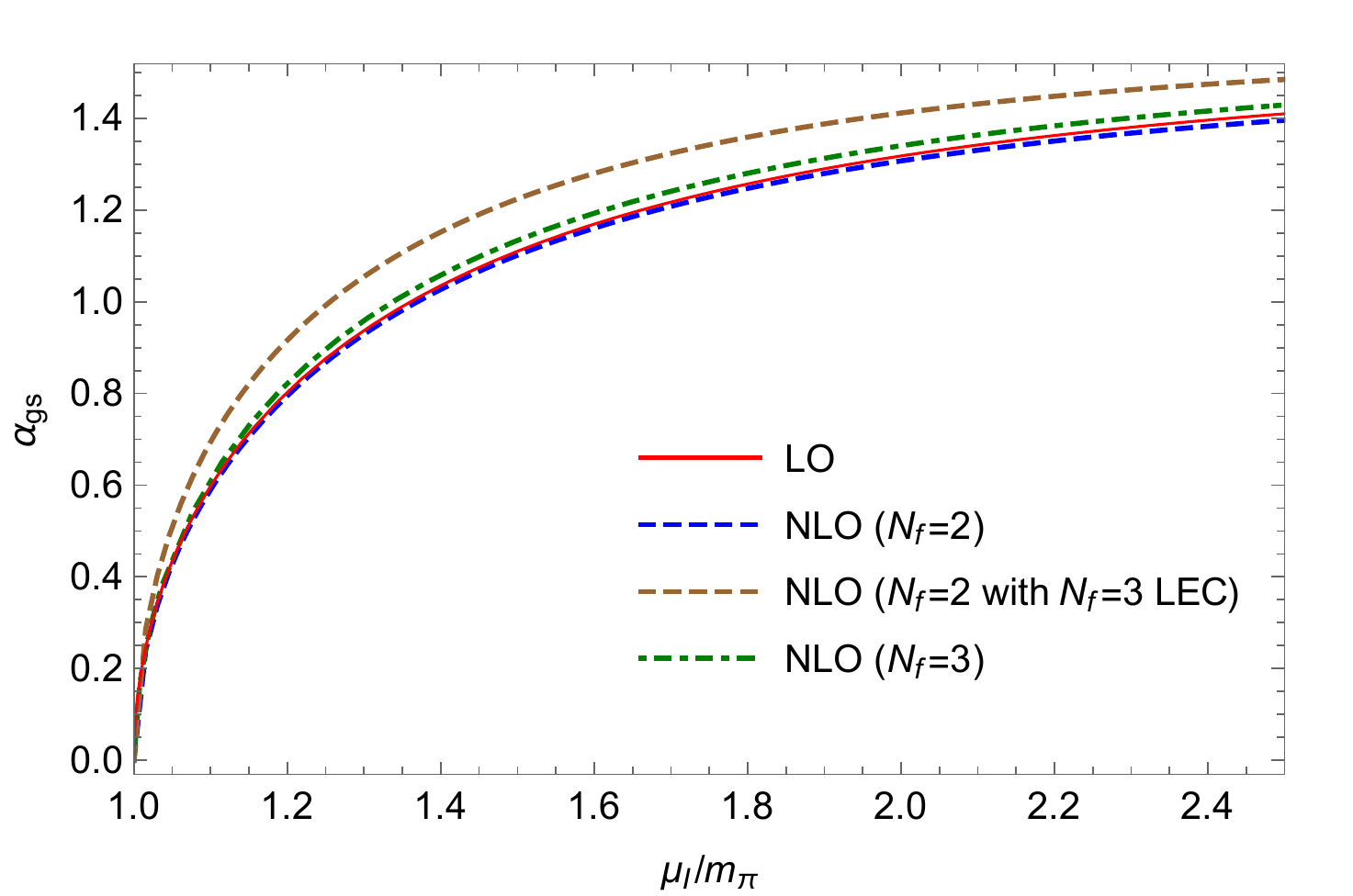}
  \caption{$\alpha_{\rm gs}$ as a function of $\mu_I/m_{\pi}$ at LO (red), at
    NLO with two flavors (blue), NLO with three flavors (green), and
   NLO with two flavors and three-flavor LECs (brown).
  See main text for details.}
\label{alphi}
\end{figure}

The thermodynamic quantities are functions of the effective potential
evaluated at its minimum as a function of $\alpha$ for given values of the
isospin and strange chemical potentials.
Hence, we must solve the equation
\bqa
{\partial V_{\text{eff}}\over\partial\alpha}&=&0\;.
\label{alphi1}
\eqa
In Fig.~\ref{alphi}, we show the solution to Eq.~(\ref{alphi1}) as
function of the isospin chemical potential $\mu_I$ and $\mu_S=0$.
The red curve is
the tree-level result, while the blue curve is the one-loop result in two-flavor $\chi$PT, the green curve is the one-loop result in three-flavor $\chi$PT and the brown curve is the one-loop result in two-flavor $\chi$PT using three-flavor LECs. In Section~\ref{pideos}, we use $\alpha_{\rm gs}$ to calculate the pressure, isospin density and the equation of state.

\subsection{Phase diagram}
\label{pd}
We find that $\alpha_{\rm gs}$ becomes non-zero when $|\mu_{I}|>m_{\pi}$. In order to show that the transition from the vacuum phase to
the Bose-condensed phase occurs at a critical chemical potential equal to the physical pion mass, we
expand the effective potential in a power series in
$\alpha$ around $\alpha=0$ up to order $\alpha^4$ to obtain an effective
Landau-Ginzburg energy functional~\cite{split2},
\bqa
V_{\rm eff}^{\rm LG}&=&a_0+a_2\alpha^2
+a_4\alpha^4+{\cal O}(\alpha^6)\;.
\eqa
%where we have indicated that the coefficients are functions of the
%two chemical potentials $\mu_I$ and $\mu_K$.
As pointed out before,
in the charged pion-condensed phase, $V_{\rm eff}$ and therefore
the coefficients are independent of $\mu_S$.
Similarly, in the charged kaon-condensed phase, they only depend on the
combination
${1\over2}\mu_I+\mu_S$, and in the neutral kaon-condensed phase, only
on the combination $-{1\over2}\mu_I+\mu_S$, 
Using the expressions for the pion mass $m_{\pi}$ 
(\ref{mpi}) and the pion-decay constant $f_{\pi}$, (\ref{fpi}),
it can be shown that in the pion-condensed phase
(see Ref.~\cite{us} for details)
\bqa
%V_{\rm eff}&=&%-f_{\pi}^2m_{\pi}^2
%V_{\rm eff}(\alpha=0)
a_2(\mu_I)&=&
{1\over2}f_{\pi}^2\left[m_{\pi}^2-\mu_I^2\right]\;.
\label{a2}
\eqa
The critical isospin chemical potential $\mu_I^c$ is defined by the vanishing
of $a_2(\mu_I)$, and Eq.~(\ref{a2}) shows that $|\mu_I^c|=m_{\pi}$.
Moreover,
using the techniques in Ref.~\cite{split2}
it can be shown that $a_4(\mu_I^c)>0$, implying that the 
the transition from the vacuum phase to a pion-condensed
phase is second order located at
$\mu_I^c=\pm m_{\pi}$.~\footnote{If $a_4(\mu_I^c)<0$, the transition is first
  order.}
Similarly, in the charged kaon-condensed phase,
we find 
\bqa
%V_{\rm eff}&=&V_{\rm eff}(\alpha=0)+{1\over2}f_K^2
a_2(\mbox{$1\over2$}\mu_I+\mu_{S})&=&\frac{1}{2}f_{K}^{2}
\left[m_K^2-\left(\mbox{$1\over2$}\mu_I+\mu_S\right)^2\right]
\;,
\eqa
where $m_K$ is the physical kaon mass, whose one-loop expression
is given by Eq.~(\ref{mk}). The critical chemical potential
is again given by the vanishing of $a_2$, i.e.
$|\mbox{$1\over2$}\mu_I+\mu_S|=m_K$
The coefficient of the order $\alpha^4$ term
can be shown to be positive when evaluated at ${1\over2}\mu_I+\mu_S=m_K$.
This shows there is a second-order transition to
a kaon-condensed phase at $\mbox{$1\over2$}\mu_I+\mu_S=\pm m_K$.
For the transition to a neutral kaon-condensed phase, we have
$-\mbox{$1\over2$}\mu_I+\mu_S=\pm m_K$.

\begin{figure}[htb]
\centering  
\includegraphics[scale=.8]{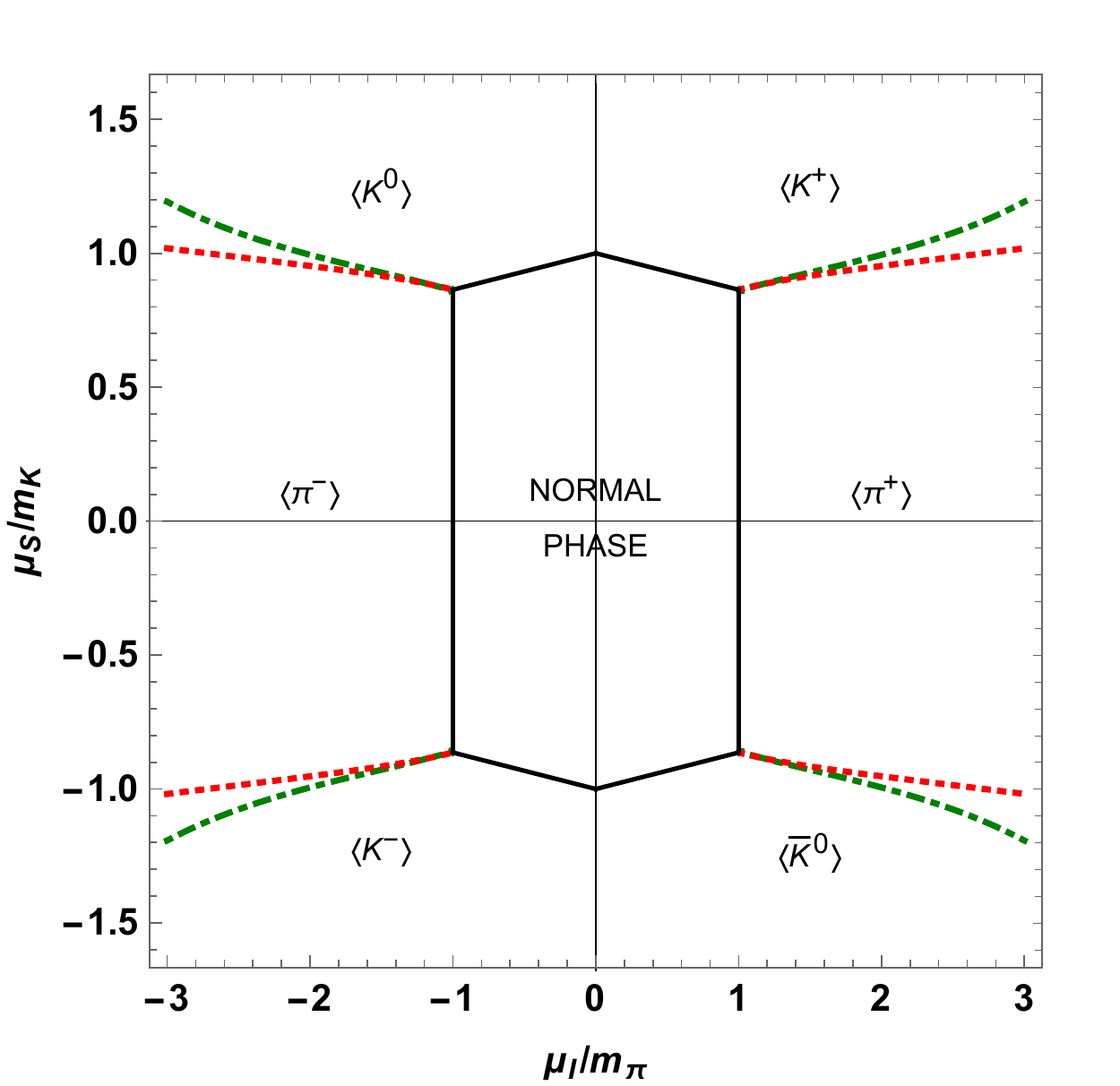}
\caption{Phase diagram in the $\mu_I$--$\mu_S$ plane at $T=0$. Solid lines
  represent second-order transitions while the red dashed lines are the tree level first-order transitions and the green dot-dashed lines are the NLO first-order transitions. The normal phase has vanishing meson condensates and the meson condensate that becomes non-zero is indicated in each region.}

\label{diagram}
\end{figure}

While the transitions from the vacuum to either a pion-condensed phase
or a kaon-condensed phase are second order, the transition between
the two Bose-condensed phases is first order.
At leading, this is straightforward to see.
For example the pion and kaon condensates are given by
\bqa
\langle\pi^+\rangle&=&2f^2B_0\sin\alpha
=2f^2B_0\sqrt{1-{m_{\pi}^4\over\mu_I^4}}\;,
\mu_I>m_{\pi}
\\
\langle K^+\rangle&=&2f^2B_0\sin\alpha
=2f^2B_0\sqrt{1-{m_{K}^4\over(\mbox{$1\over2$}\mu_I+\mu_S)^4}}
\;,\mbox{$1\over2$}\mu_I+\mu_S>m_K
\;.
\eqa
For any $\mu_I>m_{\pi}$ and $\mbox{$1\over2$}\mu_I+\mu_S>m_K$, these
condensates jump discontinuously to zero as we cross the
phase line. The transition line itself is given by the
equality of the pressures in the two phases. While it is possible to find this line analytically at tree level as shown in Eq.~(\ref{phaseline}), in order to find the line at NLO, we need to compare the pressure in the pion and kaon condensed phases, which can only be done numerically. We have performed this calculation for the central values from Eqs.~(\ref{LECs}) and (\ref{latticeval}). In Fig.~\ref{diagram} we show the phase diagram in the $\mu_I$--$\mu_S$ plane with the first order transition line increasing to higher strange chemical potential for all values of the isospin chemical potential greater than the pion mass.
The vacuum phase is in the region bounded by the straight lines
$\mu_I=\pm m_{\pi}$, $\mu_S=\pm({1\over2}\mu_I+m_K)$, and
$\mu_S=\pm(-{1\over2}\mu_I+m_K)$.
The corners from where the first-order lines emerge
are located at $(\mu_I,\mu_S)=(\pm131,\pm415.5)$ MeV.
The solid lines represent second-order transitions while the dashed
line indicates the tree level first-order transition and the green dot dashed line indicates the NLO first-order transition.
In the vacuum phase, the thermodynamic functions are independent of
the isospin and strange chemical potentials. This is an example
of the so-called Silver Blaze property~\cite{cohen}.

\subsection{Medium-dependent masses}
\begin{figure}[htb]
\includegraphics[scale=0.43]{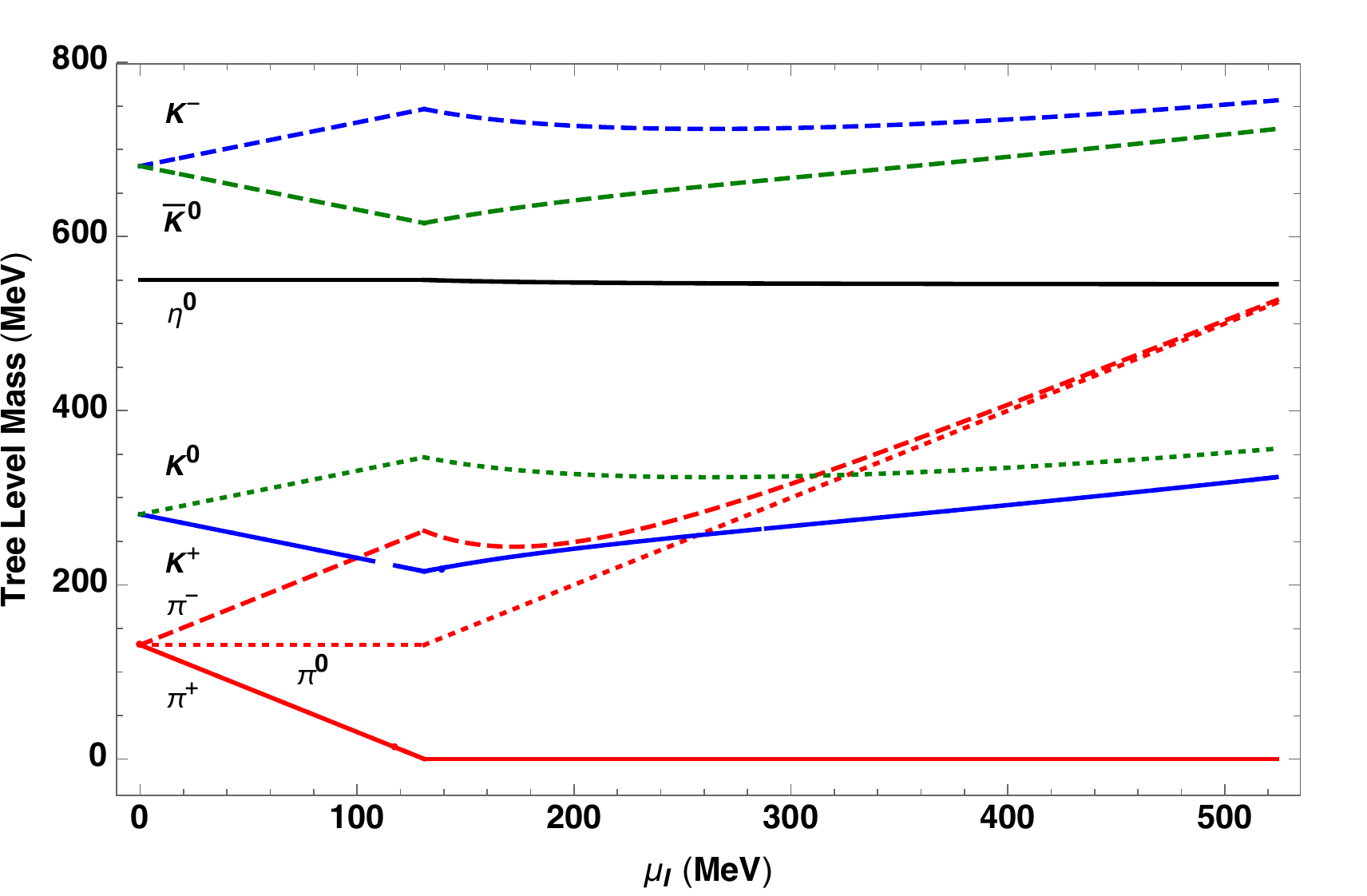}
\includegraphics[scale=0.43]{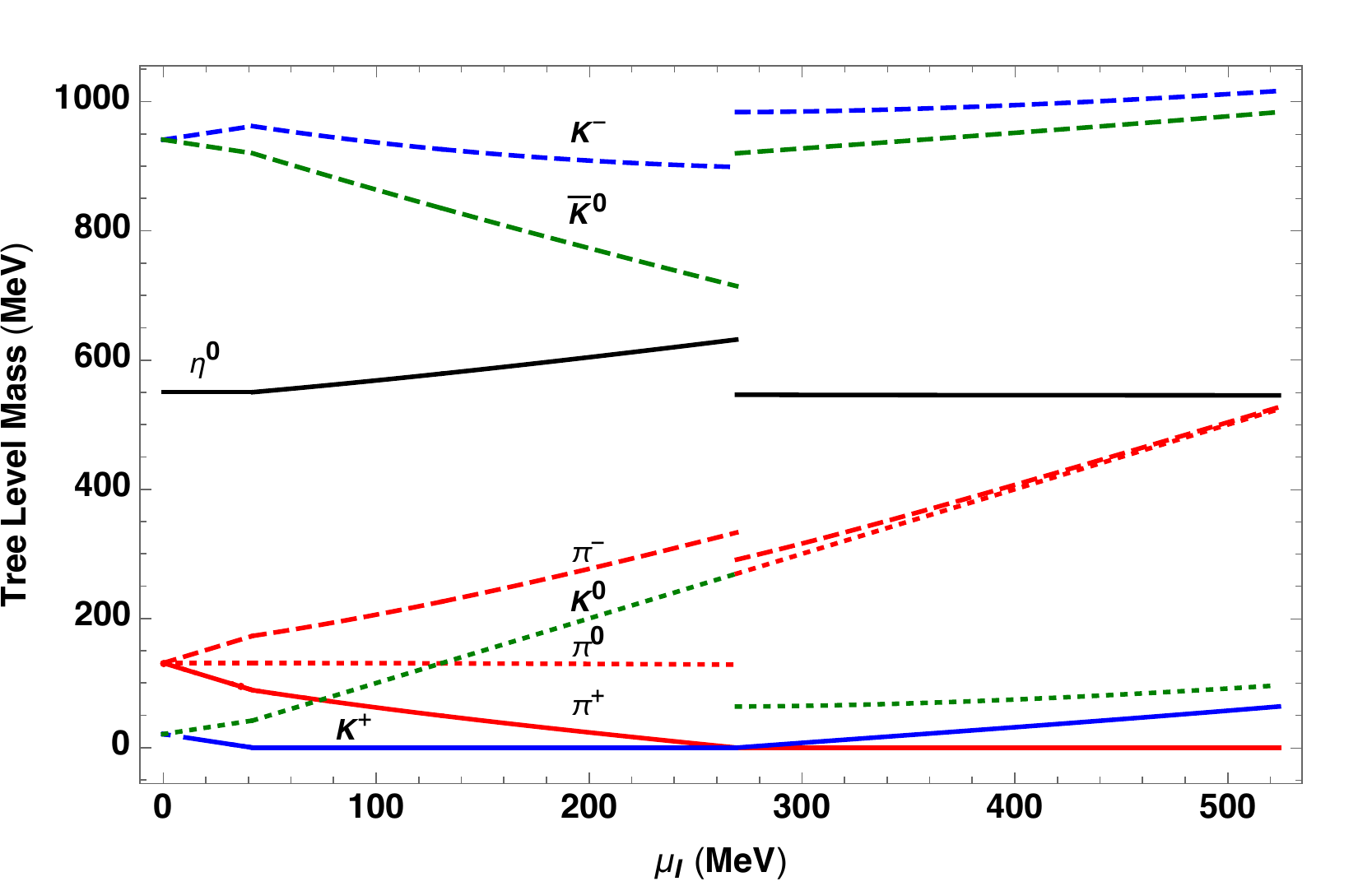}
\caption{Medium-dependent masses as a function of
  $\mu_I$ for $\mu_S=200$ MeV (left panel) and $\mu_S=460$ MeV (right panel).
See main text for details.}
\label{mediumdepmass}
\end{figure}

  In this subsection, we will briefly discuss the medium-dependent masses.
We restrict ourselves to a leading-order calculation, i.e.
we consider the tree-level dispersion relations evaluated at $p^2=0$.
In the pion-condensed phase, they are given by
Eqs.~(\ref{kvasi1})--(\ref{kvasi2}). In the kaon-condensed phase, they
are given by Eqs.~(\ref{kvasi3})--(\ref{kvasi4}).
In the left panel of Fig.~\ref{mediumdepmass}, we show the medium-dependent
masses as a function of the isospin chemical potential $\mu_I$ for
fixed strange chemical potential $\mu_S=200$ MeV.
For $\mu_I=0$, we are in the normal phase, the pion masses take on their
vacuum values, while the kaons are degenerate in pairs.
The mass of $\pi^+$ decreases as we increase $\mu_I$ and 
vanishes when $\mu_I=m_{\pi}$ and enter the pion-condensed phase.
At $\mu_I=m_{\pi}$, the masses vary continuously reflecting the second-order
nature of the transition.
We also note that the mass of $\eta^0$ is independent of
$\mu_I$, which follows directly from Eq.~(\ref{kvasi2}).
Finally, for asymptotically large values of $\mu_I$, the kaons and pions
are pairwise degenerate.
In the right panel of Fig.~\ref{mediumdepmass}, we show the medium-dependent
masses as a function of isospin
chemical potential $\mu_I$ for
fixed strange chemical potential $\mu_S=460$ MeV.
At $\mu_I=0$, we are in the vacuum phase. The kaons are again
degenerate in pairs, the pions are also degenerate taking on their
vacuum values.
We enter the kaon-condensed
phase at $\mu_I=42$ MeV, which is a second-order transition.
In this phase, $K^+$ is the Goldstone mode associated with the
spontaneous breakdown of the $U(1)$-symmetry.
As we increase the isospin chemical potential past approximately $\mu_I=268$ MeV,
we enter the pion-condensed phase. In this phase, $\pi^+$ is the
Goldstone mode associated with the
spontaneous breakdown of the $U(1)_{I_3}$-symmetry.
This first-order nature of the transition 
can be seen by the jumps in the quasiparticle masses.

Finally, we also note that in the charged pion and kaon condensed phases the mass eigenstates do not coincide with the charge eigenstates~\cite{intrigue}. It is easy to see using the form of the inverse propagators in Eqs.~(\ref{quad}) and (\ref{quad1}) that in the condensed phases the mass eigenstates can be found using momentum-dependent rotations of the mass eigenstates. However, the pion and kaon charge eigenstates are the standard ones
\begin{equation}
\begin{split}
\pi^{\pm}&=\frac{\phi_{1}\mp i\phi_{2}}{\sqrt{2}},\ K^{\pm}=\frac{\phi_{4}\mp i\phi_{5}}{\sqrt{2}}\ .
\end{split}
\end{equation}
They can be deduced using the canonical form of the quadratic, kinetic terms in Eqs.~(\ref{linear}) and (\ref{linear1}) in the unbroken phase with $\alpha=0$, which possesses a global $U(1)$ symmetry. When gauged (using electromagnetic fields), the Lagrangian possesses a local $U(1)$ (gauge) symmetry, which is broken by the pion condensed phase.

\subsection{Pressure, isospin density, and equation of state}
\label{pideos}
\begin{figure}[htb]
\centering\includegraphics[width=0.5\textwidth]{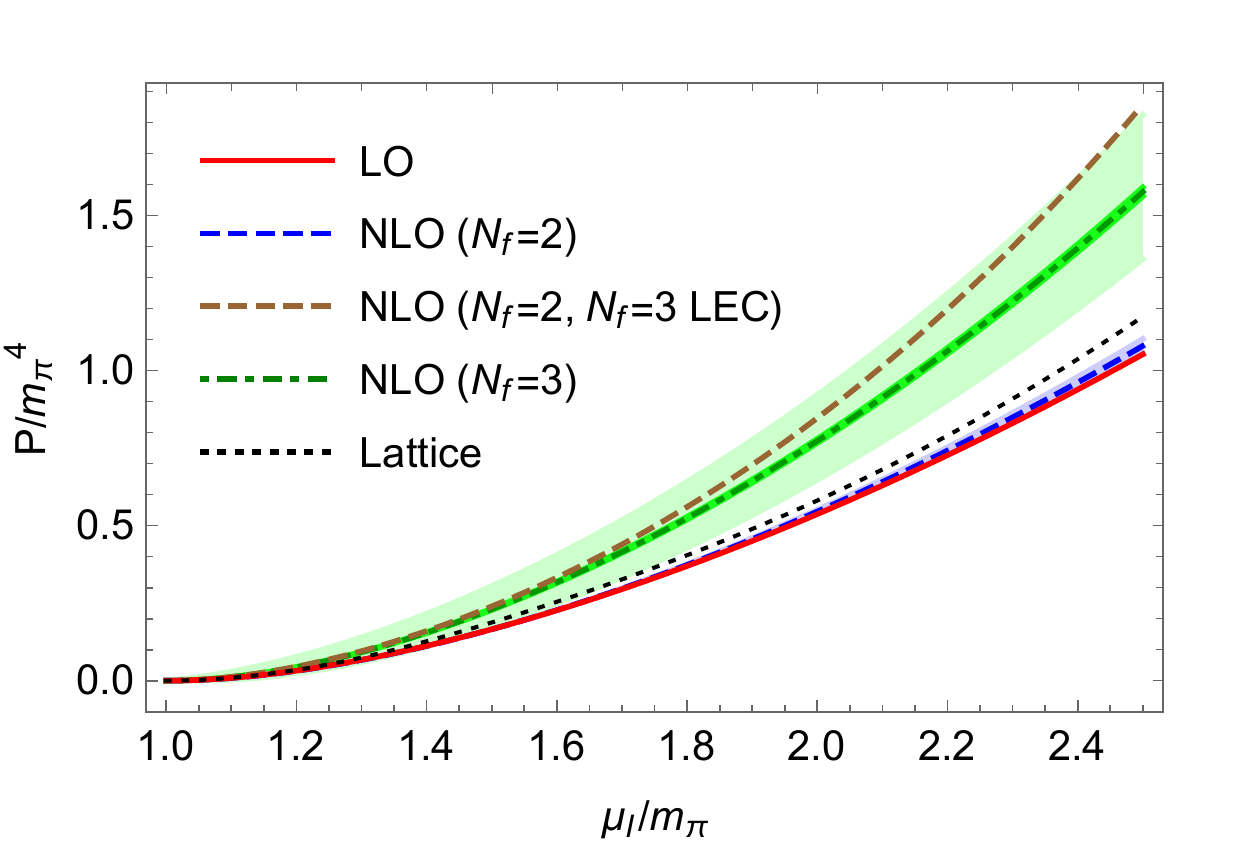}
  \caption{Normalized Pressure ($P/m_{\pi}^4$) as a function of $\mu_I/m_{\pi}$
    at LO (red), at
    NLO with two flavors (blue), NLO with three flavors (green), and 
           NLO with two flavors and three-flavor LECs (brown).
    See  main text for details.}
  \label{presieure}
  \end{figure}

In this subsection, we discuss the pressure, the isospin density and the
equation of state in the pion-condensed phase and compare our results to the
$(2+1)$-flavor lattice QCD results of Refs.~\cite{gergy1,gergy2,gergy3}. We begin
with Fig.~\ref{presieure}, where we plot the pressure (divided by $m_{\pi}^{4}$)
as a function of $\mu_{I}/m_{\pi}$. The pressure has been normalized to be zero
in the normal vacuum, which also has zero isospin density. As pions condense
beginning at the critical isospin chemical potential, $\mu_{I}^{c}=m_{\pi}$, the
pressure increases with increasing chemical potential and continues to increase
monotonically, a feature that is consistent with results from lattice QCD. The
pressure from two-flavor $\chi$PT is smaller than that from lattice QCD even
when the uncertainties within the LECs, the pion mass and pion decay constant
are taken into account. The range of pressures due to the uncertainties
calculated within two-flavor $\chi$PT is represented by the blue band. We find
that the uncertainty in the pion mass and the pion decay constant
(as opposed to the uncertainty in the LECs) dominates the uncertainty in the
pressure. On the other hand, pressure from three-flavor $\chi$PT
(shown in green), which includes the contribution from strange quarks unlike
two-flavor $\chi$PT, overestimates the pressure. In Fig.~\ref{presieure}, we
use a dark green band to show the uncertainty in the pressure due to the
uncertainties in the pion mass and the pion decay constant, and we use a light
green band to represents the uncertainty in the pressure due to the LECs, the
pion mass and the pion decay constant. The result shows that unlike in
two-flavor $\chi$PT, the uncertainty in the pressure is dominated by the
uncertainty in the LECs.
  
  It is clear from Fig.~\ref{presieure} that the difference in pressure
  calculated
  in two-flavor $\chi$PT versus that calculated in three-flavor $\chi$PT is
  quite
  significant. The tree level pressure in two and three-flavor $\chi$PT is
  identical. Therefore, the difference arises through the NLO contribution to
  the pressure in two-flavor and three-flavor $\chi$PT. Since the NLO
  contribution is suppressed by a power of $1/(4\pi f_{\pi})^{2}$, the difference
  in the two-flavor and three-flavor pressure seems unusually large. In order to
  explain the difference we have mapped three-flavor $\chi$PT by expanding the
  effective potential in the limit of large strange quark masses in
  subsection~\ref{mapping}.
  After identifying the appropriate two-flavor LECs in
  terms of three-flavor LECs -- see Eqs.~(\ref{rel0})--(\ref{rel4}) --
  we find the appropriate
  two-flavor LECs that are consistent with the values of three-flavor LECs. They
  are
\begin{equation}
\label{newLEC}
\bar{l}_{1}(N_{f}=3)=14.5,\ \bar{l}_{2}(N_{f}=3)=6.5,\ \bar{l}_{3}(N_{f}=3)=4.1,\
\bar{l}_{4}(N_{f}=3)=4.2\ ,
\end{equation}
with $\bar{l}_{i}$ being defined in Eq.~(\ref{lirlbare}). In order to contrast
the above values with the two-flavor LECs, we state the LECs below
where 
\begin{equation}
\label{LEC2old}
\bar{l}_{1}(N_{f}=2)=-0.4,\ \bar{l}_{2}(N_{f}=2)=4.3,\ \bar{l}_{3}(N_{f}=2)=2.9,
\ \bar{l}_{4}(N_{f}=2)=4.4\ .
\end{equation}
We note that while $\bar{l}_{4}$ looks quite similar in the two cases,
$\bar{l}_{2}$ and $\bar{l}_{3}$ are somewhat different with the difference in
$\bar{l}_{1}$ being the most significant (they have opposite signs).
We calculated the pressure in two-flavor $\chi$PT using the LECs found to be
consistent with three-flavor $\chi$PT -- we show this result in
Fig.~\ref{presieure} in brown (dashed).
The result shows that even the two-flavor $\chi$PT overestimates the pressure
compared to that from $2+1$ flavor lattice QCD. This analysis shows that the
overestimation of the pressure is due to the values of the LECs of three-flavor
$\chi$PT,  which also have large uncertainties compared to two-flavor LECs. As
a secondary observation, we note that as the strange quark mass becomes
lighter, the pressure increases in $\chi$PT, particularly for larger isospin
chemical potential.

\begin{figure}[htb]
\centering  \includegraphics[width=0.5\textwidth]{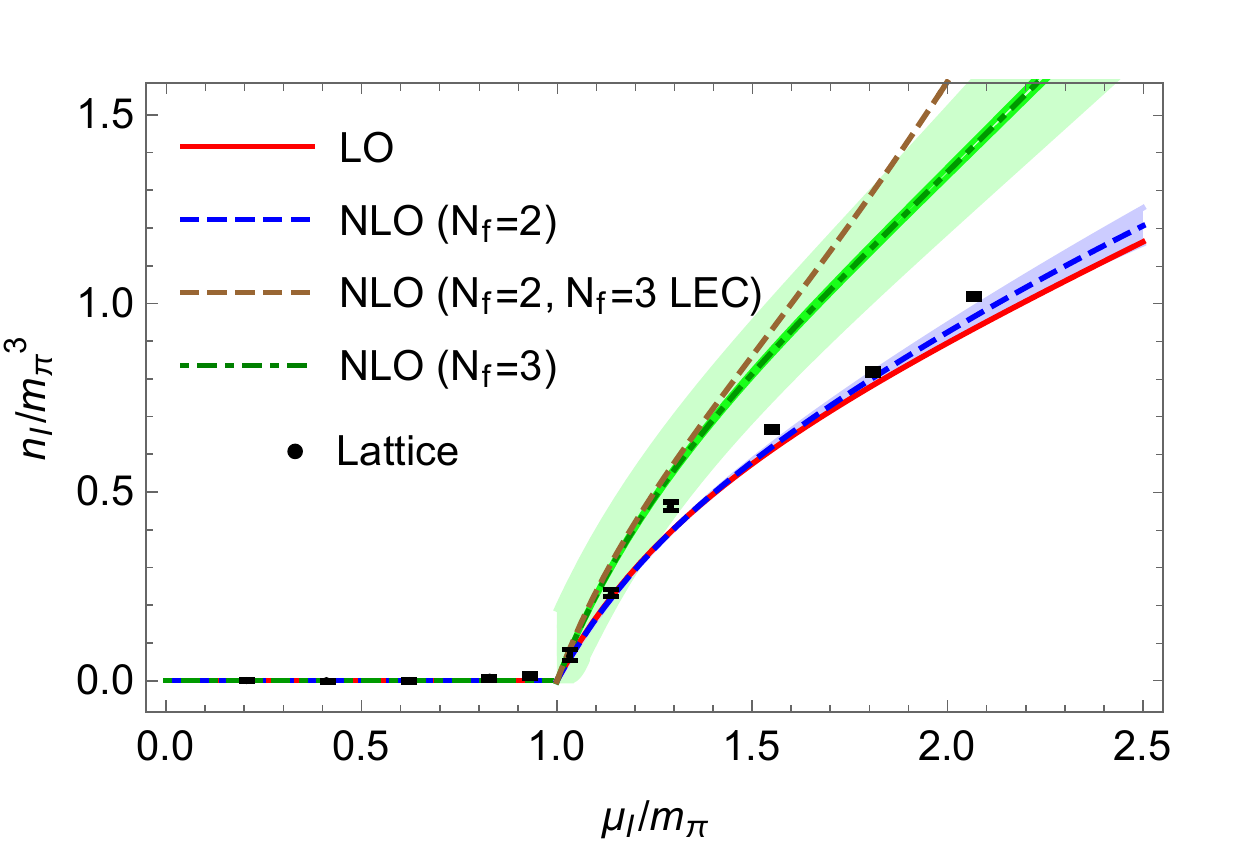}
\caption{Normalized isospin density ($n_{I}/m_{\pi}^{3}$) as a function of
  $\mu_I/m_{\pi}$
    at LO (red), at
    NLO with two flavors (blue), NLO with three flavors (green), and
       NLO with two flavors and three-flavor LECs (brown).
    See  main text for details.}
\label{iso}
\end{figure}

In Fig.~\ref{iso}, we plot the isospin density (divided by $m_{\pi}^{3}$) as a
function of the normalized chemical potential ($\mu_{I}/m_{\pi}$). The isospin
density is zero in the vacuum phase and monotonically increases in the
pion-condensed phase. The rate of increase decreases as the isospin chemical
potential increases. The isospin density from three-flavor $\chi$PT is
consistent with that of lattice QCD (in the normal vacuum and) near the
critical isospin chemical potential up to approximately $\mu_{I}=1.4m_{\pi}$. For
larger isospin chemical potentials, three-flavor $\chi$PT consistently
overestimates the isospin density. This is unlike the result in two-flavor
$\chi$PT which is in extremely good agreement with lattice QCD. The two-flavor
$\chi$PT result using three-flavor LECs is plotted in brown and shows that the
three-flavor $\chi$PT result is largely explained by the discrepancy in the
values of the LECs in two-flavor and three-flavor $\chi$PT.

\begin{figure}[htb]
\centering
  \includegraphics[width=0.5\textwidth]{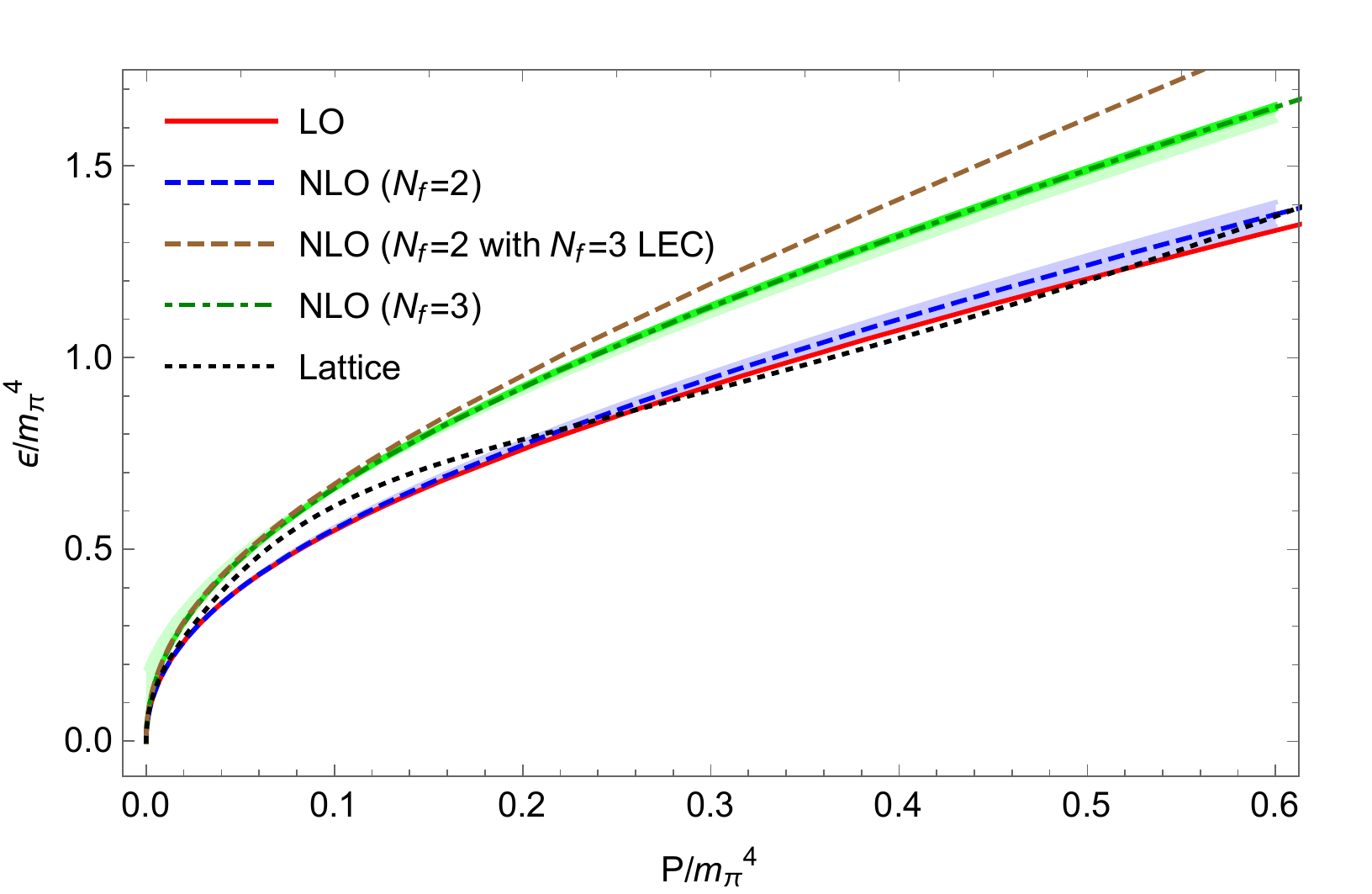}
  \caption{Normalized energy density ($\epsilon/\mu_{I}^{4}$) as a function of
    the normalized pressure ($P/m_{\pi}^4$) at
    NLO with two flavors (blue), NLO with three flavors (green), and 
           NLO with two flavors and three-flavor LECs (brown).
    See  main text for details.}
  \label{eoss}
\end{figure}

Finally, in Fig.~\ref{eoss} we plot the equation of state: the energy density
(divided by $m_{\pi}^{4}$) is plotted against the pressure
(divided by $m_{\pi}^{4}$). 
Three-flavor $\chi$PT consistently overestimates the energy density for all
pressures though up to $P/m_{\pi}^{4}\approx 0.10$, the discrepancy
is small. Two-flavor $\chi$PT, on the other hand, underestimates the energy
density
up to $P/m_{\pi}^{4}\approx0.2$ but is largely consistent for values above it.
Using three-flavor LECs, we can show that most of the discrepancy between
two-flavor and three-flavor $\chi$PT is due to the discrepancy between the two
sets of LECs, a theme common to all observables we have calculated in this work.
The two-flavor results using three-flavor LEC is shown using brown (dashed)
lines. The result is consistent with three-flavor $\chi$PT for $P/m_{\pi}^{4}\approx 0.2$ while above it the two-flavor $\chi$PT result gives larger
values of energy density. It is also worth noting that for a given value of
pressure the energy density decreases with decreasing strange quark masses.

\section{Acknowledgements}
The authors would like to thank B. Brandt, G. Endr\H{o}di
and S. Schmalzbauer
for useful discussions as well as for providing the
data points of Ref. \cite{endro}. P.A. would like to acknowledge the hospitality
of the Niels Bohr International Academy, where a portion of the work was done.
%\newpage
\appendix
%\section{Dimensionally Regularized Integrals}
%We specifically need a single integral in $d=3-2\epsilon$ dimensions,
%\bqa
%\label{int1}
%\int_p\sqrt{p^2+m^2}&=&
%-{m^4\over2(4\pi)^2}
%\left({{\Lambda^2\over m^2}}\right)^{\epsilon}
%\left[{1\over\epsilon}+{3\over2}+{\cal O}(\epsilon)\right]\;.
%\label{int2}\int_p{1\over\sqrt{p^2+m^2}}&=&-{2m^2\over(4\pi)^2}
%\left({{\Lambda^2\over m^2}}\right)^{\epsilon}
%\left[{1\over\epsilon}+1+{\cal O}(\epsilon)\right]\;,\\
%\eqa
%We need a single integral in $d=4-2\epsilon$ dimensions,
%\bqa\nonumber
%A(m^2)&=&
%\int_p{1\over p^2-m^2}\\
%&=&{im^2\over(4\pi)^2}
%\left({{\Lambda^2\over m^2}}\right)^{\epsilon}
%\left[{1\over\epsilon}+1+{\cal O}(\epsilon)
%\right]\;.
%\label{d4}
%\eqa

\section{Meson masses and decay constants}
In order to show the second-order nature of
the phase transition from the vacuum
to a Bose-condensed phase at $\mu_I^c=m_{\pi}$,
and $\pm{1\over2}\mu_I^c+\mu_S^c=m_K$
where $m_{\pi}$ and $m_K$ are 
the physical meson masses in the vacuum, we
need to express them in terms
of the parameters $B_0m$, $B_0m_s$, and $f$ of the chiral Lagrangian.
The pion and kaon masses are~\cite{gasser2}
  \bqa\nonumber
  m_{\pi}^2&=&
  m_{\pi,0}^2\left[1
    -\left(8{L}_4^r+8{L}_5^r-16{L}_6^r-16{L}_8^r
      +{1\over2(4\pi)^2}\log{\Lambda^2\over m_{\pi,0}^2}\right)
    {m_{\pi,0}^2\over f^2}
\right.\\ &&\left.
-({L}_4^r-2{L}_6^r){16m_{K,0}^2\over f^2}
  +{m_{\eta,0}^2\over6(4\pi)^2f^2}\log{\Lambda^2\over m_{\eta,0}^2}
  \right]\;,
  \label{mpi}\\ \nonumber
m_{K}^2&=&
m_{K,0}^2\left[1
  -\left({L}_4^r-2{L}_6^r\right){8m_{\pi,0}^2\over f^2}
%\right.\\ &&\left.
    -(2L_4^r+{L}_5^r-4L_6^r-2{L}_8^r){8m_{K,0}^2\over f^2}
\right.\\ &&\left.  -{m_{\eta,0}^2\over3(4\pi)^2f^2}
  \log{\Lambda^2\over m_{\eta,0}^2}
\right]\;.
\label{mk}
\eqa
%where we have used Eq.~(\ref{lowl}) with $i=3$, and Eq.~(\ref{d4})
The pion and kaon decay constants, $f_{\pi}$ and $f_{K}$ respectively, are~\cite{gasser2}
%can be determined in a similar manner at NLO,
\bqa\nonumber
\label{fpi}
f_{\pi}^2
&=&f^2\left[1
  +\left(8{L}_4^rs
+8L_5^r+{2\over(4\pi)^2}\log{\Lambda^2\over m_{\pi,0}^2}
\right){m_{\pi,0}^2\over f^2}\right.\\
&&\left.+\left(16L_4^r+{1\over(4\pi)^2}\log{\Lambda^2\over m_{K,0}^2}
  \right){m_{K,0}^2\over f^2}
\right]\\ \nonumber
f_K^2&=&f^2\left[1+
    \left(12L_4^r+
 {3\over4(4\pi)^2}\log{\Lambda^2\over m_{\pi,0}^2}\right){m_{\pi,0}^2\over f^2}
+\left(8L_5^r+
{3\over2(4\pi)^2}\log{\Lambda^2\over m_{K,0}^2}\right){m_{K,0}^2\over f^2}
\right.
\\ &&\left.
  +\left(12L_4^r+
    {3\over4(4\pi)^2}\log{\Lambda^2\over m_{\eta,0}^2}\right)
{m_{\eta,0}^2\over f^2}\right]
\;.
\label{fk}
\eqa
Using the expressions for the renormalization group
equations, Eq.~(\ref{rgeq0}), it is
straightforward to see that the $\Lambda$-dependence of the coupling
cancels against the chiral logarithms in expressions for the
masses and decay constants.
  
\bibliographystyle{elsarticle-num} 
\bibliography{bib}

\end{document}